\documentclass[prd,aps,
nofootinbib,
floatfix,
superscriptaddress,preprintnumbers]{revtex4-1}
\usepackage{graphicx,epstopdf}
\usepackage{epsfig}
\usepackage{rotating}
\usepackage{amssymb}
\usepackage{subfigure}
\usepackage{dsfont}
\usepackage{psfrag}
\usepackage{amsmath,euscript,array,mathrsfs}
\usepackage{axodraw}
\usepackage{bbold}

\newcommand{\beq}{\begin{equation}}
\newcommand{\eeq}{\end{equation}}
\newcommand{\beqs}{\begin{eqnarray}}
\newcommand{\eeqs}{\end{eqnarray}}
\newcommand{\lsim}{\mathrel{\raisebox{-
.6ex}{$\stackrel{\textstyle<}{\sim}$}}}

\def\hbar{\hspace{0pt}\raisebox{1pt}{$-$} \hspace{-7pt} h}

\def\di{\mbox{d}}
\def\r{\rho}

\newcommand{\be}{\begin{equation}}
\newcommand{\ee}{\end{equation}}
\newcommand{\bea}{\begin{eqnarray}}
\newcommand{\eea}{\end{eqnarray}}

\def\lbldef#1#2{\expandafter\gdef\csname #1\endcsname {#2}}

\def\href#1#2{#2}

\newcommand{\ber}{\begin{eqnarray}}
\newcommand{\eer}{\end{eqnarray}}

\newcommand{\beqar}{\begin{eqnarray}}

\newcommand{\eeqar}{\end{eqnarray}}

\newcommand{\dsl}
  {\kern.06em\hbox{\raise.15ex\hbox{$/$}\kern-.56em\hbox{$\partial$}}}

\newcommand{\eeqarr}{\end{eqnarray}}
\newcommand{\ZZ}{{\rm \kern 0.275em Z \kern -0.92em Z}\;}

\def\CC{{\mathchoice
{\rm C\mkern-8mu\vrule height1.45ex depth-.05ex
width.05em\mkern9mu\kern-.05em}
{\rm C\mkern-8mu\vrule height1.45ex depth-.05ex
width.05em\mkern9mu\kern-.05em}
{\rm C\mkern-8mu\vrule height1ex depth-.07ex
width.035em\mkern9mu\kern-.035em}
{\rm C\mkern-8mu\vrule height.65ex depth-.1ex
width.025em\mkern8mu\kern-.025em}}}

\def\RR{{\rm I\kern-1.6pt {\rm R}}}

\def\ZZ{{\rm Z}\kern-3.8pt {\rm Z} \kern2pt}
\def\IB{\relax{\rm I\kern-.18em B}}
\def\ID{\relax{\rm I\kern-.18em D}}
\def\II{\relax{\rm I\kern-.18em I}}
\def\IP{\relax{\rm I\kern-.18em P}}

\newcommand{\bear}{\begin{eqnarray}}
\newcommand{\eear}{\end{eqnarray}}

\newcommand{\F}{{\cal F}}

\def\r{\rho}                                     
\def\t{\tau}

\def\6{\partial}

\def\bea{\begin{eqnarray}}
\def\eea{\end{eqnarray}}

\def\beqx{\begin{displaymath}}
\def\eeqx{\end{displaymath}}

\newcommand{\bmat}{\left(\begin{array}}
\newcommand{\emat}{\end{array}\right)}

\def\r{\rho}

\def\t{\tau}

\def\F{\Phi}

\def\bo{{\raise-.3ex\hbox{\large$\Box$}}}               
\def\face{{\raise.2ex\hbox{$\displaystyle \bigodot$}\mskip-2.2mu \llap {$\ddot
        \smile$}}}                                   
\def\>{\rangle}                                      
\def\<{\langle}                                      

\def\leftrightarrowfill{$\mathsurround=0pt \mathord\leftarrow \mkern-6mu
        \cleaders\hbox{$\mkern-2mu \mathord- \mkern-2mu$}\hfill
        \mkern-6mu \mathord\rightarrow$}        
\def\dvec#1{\vbox{\ialign{##\crcr
        \leftrightarrowfill\crcr\noalign{\kern-1pt\nointerlineskip}
        $\hfil\displaystyle{#1}\hfil$\crcr}}}           

\def\-{\hphantom{-}}

\begin{document}
\title{On the stability of multiscale models of dynamical symmetry breaking from holography.}

\author{ Anton F. Faedo}
\affiliation{Department of Physics, College of Science, Swansea University,
Singleton Park, Swansea, Wales, UK.}
\affiliation{Departament de F\'{\i}sica Fonamental  \& Institut de Ci\`encies del Cosmos, Universitat de Barcelona, Mart\'{\i}  i Franqu\`es 1, E-08028 Barcelona, Spain}
\author{ Maurizio Piai}
\affiliation{Department of Physics, College of Science, Swansea University,
Singleton Park, Swansea, Wales, UK.}
\author{Daniel Schofield}
\affiliation{Department of Physics, College of Science, Swansea University,
Singleton Park, Swansea, Wales, UK.}

\preprint{ICCUB-13-246}


\begin{abstract}
We consider two classes of backgrounds
of Type IIB supergravity obtained by wrapping $D5$-branes on  a two-cycle inside the conifold.
The field theory dual exhibits confinement and, in addition, a region in which the dynamics is walking, at least 
in the weak sense that the running of the coupling is anomalously slow. 
We introduce quenched matter in the fundamental, modelled by probe 
$D7$-branes which wrap an internal three-dimensional manifold and lie at the equator 
of the transverse two-sphere. 
In the space spanned by the remaining internal angle and the radial coordinate the branes admit two embeddings. 
The first one is U-shaped: the branes merge at some finite value of the radius. 
The second one is disconnected and extends along the entire radial direction at fixed angular separation. 
We interpret these two configurations as corresponding to chiral-symmetry breaking 
and preserving phases, respectively. 

We present a simple diagnostic tool to examine the classical stability of the embedding,
based on the concavity/convexity conditions for the relevant thermodynamic potentials. 
We use this criterion to show that U-shaped probes that explore the walking region are unstable,
hence providing a dynamical origin for the tachyonic mode found
in the literature.  Whenever this occurs, the disconnected solution becomes favored 
energetically. We find that in one of the two classes of backgrounds the U-shaped embedding is
always unstable, and thus never realised dynamically. Consequently, these models cannot be 
used to describe chiral-symmetry breaking.
In the second category of solutions, our analysis reveals the presence of a first-order
phase transition between chiral-symmetry broken  and restored phases. 
Interestingly, this is in the same class that contains 
a parametrically light scalar in the spectrum of glueballs of the dual field theory.

\end{abstract}

\maketitle

\tableofcontents

\section{Introduction}

The construction of viable (realistic and calculable) models of dynamical electroweak
symmetry breaking, usually referred to as technicolor (TC)~\cite{TC,reviewsTC},
is a notoriously difficult and challenging task.
Luckily, nowadays we have a huge body of experimental measurements guiding this process,
thanks to precision electroweak tests, to high precision 
data collected about rare decays mediated by flavor-changing neutral currents,
and to the fact that
the LHC experiments ATLAS~\cite{ATLAS} and CMS~\cite{CMS}
announced in 2012 the discovery of a new scalar particle with mass in the range of 125-126 GeV.
This whole body of evidence suggests that if electroweak symmetry breaking is due 
to a new strongly-coupled interaction, then the fundamental TC model, and its embedding in a theory of flavor
via extended technicolor (ETC)~\cite{ETC}, must have some very special, unusual properties.
In particular, the strongly-coupled model cannot be some simple generalization of a QCD-like theory,
because it must explain the large hierarchies of scales visible in experimentally accessible observables. 
This suggests that the nature of the fundamental theory of electroweak symmetry breaking should
 itself generate several parametrically separated dynamical scales.

Walking technicolor (WTC)~\cite{WTC} is such a special possibility.
The fundamental reason why models of this class are radically different from 
models inspired by QCD is that the dynamics is intrinsically multi-scale: within the range of energies
above the electroweak scale $\Lambda_W$ and below a new dynamical scale $\Lambda_{\ast}$ the theory is strongly coupled but approximately scale invariant.
As a consequence, large anomalous dimensions arise naturally, addressing the flavor problem(s),
while the presence of parametrically separated scales may be used
to soften the problems with electroweak precision tests such as the
$S$ and $T$ parameters~\cite{Peskin,Barbieri}.
A semi-realistic model is, for instance, developed and studied in~\cite{APS},
making also use of important elements from~\cite{preAPS}.
Furthermore, already in the early papers on the subject~\cite{Yamawaki} it was suggested that such models
might contain an anomalously light scalar particle in their spectrum.
This particle  is usually referred to as {\it dilaton}, to stress the fact that its comparatively light mass
and the special properties of its leading-order couplings can be explained in terms of
the spontaneous breaking of scale invariance.

Field-theoretical and phenomenological studies of the dilaton are the subject of a vast
literature~\cite{HT,dilatonpheno,dilaton4,dilatonnew,dilatonandpheno,dilaton5D}.
The general consensus is that a clear systematic understanding of what specific models give rise to a light dilaton is still under development.
Nevertheless, such a particle might coincide with the Higgs resonance discovered at the LHC,
because the main properties of the Higgs particle are due to 
the fact that it is itself a dilaton (although elementary, and arising from a weakly-coupled 
theory of electroweak symmetry breaking).
Due to the intrinsic conceptual and technical limitations 
 of analytical field-theory tools applied to such a non-trivial strongly-coupled system, 
other techniques, more suited to study 
non-perturbative physics, are needed.
On the numerical side, lattice studies made significant progress in recent years (see for example~\cite{lattice}).

On the analytical side, a powerful tool for studying strongly-coupled field theories 
makes use of  gauge/gravity dualities, arising in the
context of string theory~\cite{AdSCFT,reviewAdSCFT}.
This allows to examine non-trivial phenomena such as confinement and chiral symmetry breaking.
The first step in this direction requires to find classical solutions in the low energy (supergravity) 
limit of string or M-theory. The ansatz for the metric is a (warped) product of
an internal compact five-dimensional manifold and a non-compact five-dimensional space.
Four of the non-compact dimensions $x^{\mu}$ are identified with the Minkowski space.
The fifth non-compact dimension $\r$ is related to the renormalization scale of the dual field theory.
Such a background can describe a confining field theory provided the geometry
closes smoothly at some finite value $\r_0$ of the radial direction. 
The bulk dynamics is controlled by a set of fields that are allowed to propagate in the ten-dimensional
space. The field theory data is then recovered by 
assuming that the boundary values of the bulk fields (at $\r\rightarrow +\infty$)  
act as sources (or VEVs) of local operators. 
In this sense, the dual field theory lives at the UV boundary of the non-compact space.
One can use this setup to compute correlation functions by implementing holographic renormalization~\cite{HR}.

This procedure can be generalized to the study of non-local operators in the field theory
by considering the bulk dynamics of extended objects, usually treated in probe approximation. Wilson loops in the gauge theory can be examined by allowing a probe open string,
with end-points on a $Dp$-brane at the UV boundary, to explore the bulk geometry.
In this way, one can recover highly non-trivial results, such as the static quark-antiquark potential
both in the case of conformal~\cite{MaldacenaWilson} and confining~\cite{Witten} field theories.
Chiral symmetry breaking can be studied in a somewhat similar manner 
by letting a stack of $N_f$  $Dp$-branes probe the geometry~\cite{Karch:2002sh}. One looks for configurations
that admit classically stable U-shaped embeddings such that the probes extend from the UV boundary down to some 
finite value of the radial direction $\hat{\r}_o>\r_0$, and then turn back towards the UV.
The theory living on the stack of branes has naturally a $U(N_f)$ symmetry, but since the U-shape 
embedding is a double covering of the radial direction, effectively one finds
a $U(N_f)_L\times U(N_f)_R$ symmetry. However, given that the two branches of the embedding merge smoothly at $\hat{\r}_o$,
the dynamics realizes linearly only the diagonal $U(N_f)_V$. The result is a strongly coupled model in which the breaking $U(N_f)_L\times U(N_f)_R\rightarrow U(N_f)_V$ is taking place.
The turning point $\hat{\r}_o$ of the embedding sets the scale of chiral symmetry breaking.
This idea has been proposed and  successfully studied in~\cite{SS}, who considered the Type IIA background
of~\cite{Witten}, allowing $D8$ branes to extend in the Minkowski directions
and wrap an internal $S^4$. It was found that a U-shaped embedding exists in the remaining
two-dimensional surface described by the radial direction together with an internal $S^1$.
 
The ultimate goal of this program would be to understand the dynamics of 
QCD and QCD-like models. Nevertheless, it is very natural to use the same tools with other strongly coupled, confining theories that undergo the phenomenon of chiral symmetry breaking.
The obvious application is technicolor. Early attempts at generalizing the procedure of
Sakai and Sugimoto to various different contexts focused on the precision physics observables
of the dual technicolor theory~\cite{stringS}. Unfortunately, they were met with somewhat disappointing results: while the procedure works,
and overall the phenomenology is qualitatively as expected, precision electroweak parameters, in particular $S$,
tend to be too big for a realistic model of the electroweak theory.

The problem with the models in~\cite{stringS} can be traced back to the fact that the 
backgrounds considered are the duals of theories rather similar to QCD.
In particular, there is only one dynamical scale. Seen under this light, the results of ~\cite{stringS} yield a comforting
assessment of the whole holographic approach: the dual gravity description of
technicolor models in which all the non-perturbative phenomena are controlled by one  
dynamically generated scale gives large results for precision electroweak observables,
in agreement with the field theory expectations.
As we already stated, in order for electroweak precision tests to be met
successfully one needs a non-trivial technicolor model in which the dynamics is intrinsically multi-scale.
The next logical step is hence to look for (super)gravity backgrounds
whose dual exhibits the crucial  multi-scale nature required by a realistic technicolor model,
and then repeat the exercise of Sakai and Sugimoto by finding appropriate embeddings
of $Dp$ branes in these new backgrounds.

Starting with~\cite{NPP}, the search for such multi-scale supergravity backgrounds has to large extent been focused
on the conifold and its variations~\cite{CO}.
This is a particularly appealing context, because many well studied supergravity solutions
have been found within this framework~\cite{KW,KT,KS,MN,BGMPZ}, which is well understood. Indeed, all these
backgrounds are different solutions to the equations of Type IIB within the Papadopoulos--Tseytlin ansatz~\cite{PT}, which recently has been shown to be a subtruncation of a more general 
supersymmetric consistent truncation on $T^{1,1}$, the base of the conifold~\cite{consistentconifold}.

In~\cite{NPP}, the first class of solutions of direct relevance to the present paper was
found, by making use of the tools developed in~\cite{HNP}.
These solutions exhibit, besides the confinement scale $\r_0$ at the end-of-space of the geometry, 
a second, parametrically larger dynamical scale $\r_{\ast}>\r_0$.
With some abuse of language, solutions of this type will, in the following, be referred to as walking,
because, by adopting a specific definition for the dual gauge coupling 
borrowed from~\cite{gauge}, the resulting running is very slow over a 
finite energy interval below the scale corresponding to $\r_{\ast}$.
Subsequently, several different but related classes of walking solutions were found~\cite{ENP,NPR,EGNP}. 
It has been shown that confinement can be described in the 
familiar way also for walking solutions~\cite{NPR}, although a highly non-trivial phenomenon similar to 
a phase transition takes place in the presence of a walking region.
It has been found that an anomalously light scalar is present in the spectrum of some of these models~\cite{ENP,EP} by studying the fluctuations  of the truncated five-dimensional sigma-model
with the formalism developed in~\cite{BHMEP}.
Finally, there are backgrounds of this type not
only in the restricted context of the wrapped-$D5$ system as in~\cite{NPP}, but also
in the generalizations of the Klebanov-Strassler system to the baryonic branch~\cite{EGNP}. 
This made it possible to perform a sensible field-theory analysis 
by using the results, ideas and techniques in~\cite{quivers, rotation,GMNP,CNP-Z,dimensions}.

These developments did not go unnoticed to~\cite{A}, who reopened the case for computing the $S$-parameter 
in the supergravity dual of a multi-scale dynamical model.
The background considered in~\cite{A} is the one in~\cite{NPP}. The proposal is to use a specific 
embedding of probe $D7$ branes that wrap an internal three-dimensional manifold. 
Interestingly, in~\cite{ASW1} it was found that models of electroweak symmetry breaking based on this construction
admit regions of parameter space in which the $S$ parameter is finite, positive and small. Furthermore, smallness of $S$ is related to the separation between the scales responsible for confinement and chiral symmetry breaking.
The idea that chiral symmetry breaking could happen at a scale that does not coincide with the scale of confinement is a comparatively old one~\cite{MG} and has been explored for many different reasons. If it were true that 
we have now a concrete realization of this idea in the context of gauge/gravity dualities, and that it leads to a suppression of 
precision electroweak observables, this in turns would be a very major conceptual and practical breakthrough.

The embedding suggested in~\cite{A} is however not the result of a systematic study of
what type of configurations are admitted by the background. In its beautiful simplicity,
it leaves open the question of whether such embedding is actually stable.
Such question has been addressed in~\cite{ASW2,CLV}, where the spectrum of 
fluctuations has been studied.
The result is that while most of the fluctuating modes are healthy, there is a 
tachyon in the spectrum~\cite{CLV}, hence signalling a pathology in the system
 (see~\cite{ASW3}
for a critical discussion of the nature of such state).

The first result of this paper is a simple and elegant way of assessing wether a given embedding
is perturbatively stable or not, without having to perform the heavy task 
of explicitly computing the spectrum of fluctuations as in~\cite{ASW2,CLV,ASW3} (for another work along the same direction see~\cite{ASW4}). This criterium can be obtained as the consequence of a concavity condition similar to the ones encountered for thermodynamic potentials, as we argue using a convenient analogy. When applied to the conifold backgrounds of interest, our diagnostic tool uncovers an instability as soon as the U-shaped probes wander through the walking region. Pushing further the thermodynamic analogy, one must
 wonder about the fate of the system as it reaches the instability and, in particular, if there is a different configuration of branes that takes over the U-shaped one as it becomes unstable. 

Indeed, we will see that there is an alternative solution in which the branes extend 
along the entire radial direction and lie at a fixed angular separation in the 
transverse space. Since, contrary to the U-shaped case, the branes do not 
merge, this other arrangement is expected to preserve chiral symmetry. 
Using energetic arguments, we will show that this disconnected configuration 
is preferred whenever the connected one becomes unstable. 
For the geometries probed in~\cite{A} this gives a natural dynamical origin 
to the tachyon in the spectrum, as the brane configuration considered is 
not a minimum of the action. Furthermore, for a different family of backgrounds
 that we will detail, one comes upon a first order transition between 
 chiral-symmetry breaking and preserving phases. The control parameter
  is the asymptotic angular separation of the branes, which is expected to 
  characterize how the fundamental matter is coupled to the adjoint content already present.  

The paper is organized as follows.
In Section~\ref{Sec:conifold} we summarize the 
main features of a large class of solutions to the BPS equations
describing $D5$-branes wrapping a two-cycle inside the conifold.
These solutions will be the subject of the rest of the paper.
In Section~\ref{Sec:extended} we introduce the general formalism used to solve the equations of motion, in probe approximation, of 
an extended object living in a fixed background geometry.
We summarize a few useful results and, using a thermodynamic analogy, derive a neat criterion for stability of a 
given embedding.
 Section~\ref{Sec:applications} is the main body of the paper: we apply this criterion to the $D7$-brane embedding proposed in~\cite{A},
considering the backgrounds of Section~\ref{Sec:conifold}. We find that U-shaped branes probing the walking region are unstable, and propose, on energetic grounds, that a different, disconnected configuration takes their place.
In Section~\ref{Sec:conclusions} we discuss
our main results, compare them to the literature, and highlight further possible lines of inquiry.

\section{A class of solutions to Type IIB.}
\label{Sec:conifold}

In this section we review and summarise results that have been derived elsewhere,
classifying and reorganizing a large class of solutions to the BPS equations of 
the reduction of Type IIB supergravity on $T^{1,1}$ (the base of the conifold).
In the process, we fix the notation used throughout the paper and clarify the geometric differences between the various classes of
backgrounds of interest.

\subsection{The wrapped-D5 system.}

All the solutions that will be discussed can be obtained from what is referred to as the wrapped-$D5$ system. This is the 
geometry produced by the strong-coupling limit of a stack of $N_c$ $D5$-branes wrapping
an $S^2$ inside  $T^{1,1}$ and extending in the 
Minkowski directions $x^{\mu}$ located at the tip of the conifold~\cite{MN,HNP}.
The system we start with is a truncation of Type IIB supergravity which  includes only gravity, the dilaton $\Phi$ and the RR three-form $F_3$. We define the following vielbein:
\beqs
e_1&=&-\sin \theta\, \di \phi\,,\\
e_2&=& \di \theta\,,\\
e_3 &=&\cos \psi\, \sin\tilde{\theta}\,\di \tilde{\phi}-\sin \psi\, \di \tilde{\theta}\,,\\
e_4 &=&\sin \psi \,\sin\tilde{\theta}\,\di \tilde{\phi}+\cos \psi \,\di \tilde{\theta}\,,\\
e_5 &=&\di \psi +\cos{\theta}\,\di{\phi}+\cos\tilde{\theta}\,\di\tilde{\phi} \,,
\eeqs
where the range of the five angles spanning the internal space  is
$0\leq \theta\,,\,\tilde{\theta}<\pi\,,\,
0\leq\phi\,,\,\tilde{\phi}<2\pi\,,\,0\leq\psi<4\pi$. We assume that the functions appearing in the background depend only the radial coordinate $\r$. The ansatz for the metric in Einstein frame takes the form
\bea
\di s^2_E &=& \alpha' g_s \,e^{\Phi/2} \Big[ (\alpha' g_s)^{-1} dx_{1,3}^2 + ds_6^2 \Big], \nonumber\\[3mm]
\di s_6^2 &=& e^{2k}d\rho^2
+ e^{2 h}
(e_1^2 + e_2^2) +
\frac{e^{2 {g}}}{4}
\left(\frac{}{}(e_3+a\,e_1)^2
+ (e_4+a\,e_2)^2\right)
+ \frac{e^{2 k}}{4}e_5^2\,.
\label{nonabmetric424}
\eea
When $a=0$, it is easy to identify the internal metric as a U(1) fibration (with fiber coordinate $\psi$ in $e_5$) over ${\rm S}^2\times {\rm S}^2$. Notice that in addition, the difference of warp factors $g-h$ breaks the $\mathbb{Z}_2$ symmetry that interchanges the spheres. 

This metric is supported by a non-vanishing $F_3$, the particular form of which will be of no use in this paper and can be found for instance in~\cite{HNP}. The full background is then determined by solving the BPS equations for the warp factors and the dilaton as a function of the radial direction $\r$. 
From here on we set $\alpha^{\prime}g_s=1$ and as usual the string-frame metric is given by $\di s^2 = e^{\frac{\Phi}{2}}\di s^2_E$.

The system of equations derived using this ansatz can be rearranged in terms of a more convenient set of functions as in~\cite{HNP}:
\beqs
4 \,e^{2h}=\frac{P^2-Q^2}{P\cosh\tau -Q}\,, \qquad\qquad e^{2{g}}= P\cosh\tau -Q,\qquad\qquad
e^{2k}= 4 \,Y\,,\qquad\qquad a=\frac{P\sinh\tau}{P\cosh\tau -Q}\,.
\label{functions}
\eeqs
Various combinations of the BPS equations can be integrated in closed form, so that the system reduces to a 
single decoupled second-order equation for the function $P(\rho)$ that reads
\beq
P'' + P'\,\Big(\frac{P'+Q'}{P-Q} +\frac{P'-Q'}{P+Q} - 4 
\coth(2\rho-2{\rho}_0)
\Big)=0.
\label{Eq:master}
\eeq
The rest of the functions are obtained from it as follows:
\begin{equation}
\begin{array}{rclcrcl}
Q&=&(Q_0+ N_c)\cosh\tau + N_c \,(2\rho \cosh\tau -1)\,,&\qquad\qquad\qquad&Y&=&\frac{P'}{8}\,,\\[3mm]
e^{4\Phi}&=&\frac{e^{4\Phi_0} \cosh(2{\rho_0})^2}{(P^2-Q^2) Y\sinh^2\tau}\,,&\qquad\qquad\qquad& \cosh\t&=&\coth(2\r-2{\r_0})\,.
\end{array}
\label{BPSeqs}
\end{equation}

We will refer to Eq.~(\ref{Eq:master}) as the {\it master equation}: this is the only 
non-trivial differential equation that needs to be solved in order to 
generate the large class of solutions
we are interested in. We will always take the end of space at $\r_0=0$, which amounts 
to setting to unity the dynamical scale 
in terms of which all other dimensionful parameters will be measured.
Also, in order to avoid a nasty singularity in the IR we fine-tune $Q_0=-N_c$.

From the set of integration constants, we have adjusted $\r_0$ and $Q_0$ to special choices. 
We could also remove a third one from the dilaton, $\Phi_0$,
which can always be reabsorbed into the definition of $\alpha' g_s$. 
For later convenience, we do no perform this rescaling: the general solution will depend explicitly on
$\Phi_0$, in spite of the fact that solutions that differ only by this parameter are
(at the semi-classical level) dynamically equivalent.
The final two integration constants appear in the solutions to the master 
equation, for which we require only that
the function $P$ be regular for any $\r\geq 0$, 
hence defining a two-parameter class of solutions to the wrapped-$D5$ system.

\subsection{Classification of possible solutions.}

The master equation is non-linear and, even setting $Q_0=-N_c$, it has an inherent tendency towards
producing bad singularities (see the denominators in the equation itself).
Besides, smoothness of $P$ does not ensure that the background is 
free of singularities, because at the end of space in the IR ($\r\rightarrow 0$)
some of the functions in the ten-dimensional ansatz may diverge or vanish.
Hence, looking for regular solutions is an intricate task.

Nevertheless, it turns out that all the possible solutions for which the function $P$ is regular
can be very roughly approximated by the following expression
\beqs\label{approxsol}
P&\simeq&P_a\,\equiv\,{\rm sup}\left\{c_0\,,\,2N_c \r\,,\,3c_+e^{\frac{4\r}{3}}\right\}\,,
\eeqs
where $c_+\geq 0$ and $c_0\geq 0$ are two integration constants.
One reason why the approximation is rough is that the actual solutions are 
smooth functions, while Eq.~(\ref{approxsol}) is not differentiable.
Yet, it serves for illustrational purposes in the context of this paper.
Effectively, $P_a$ is constructed in such a way as to ensure that both $P$ and $P^{\prime}$
be monotonically non-decreasing, and that for any $\r\geq 0$ one has $P>Q$. 
The combination of these requirements
ensures that the solution for $P$ is well-defined everywhere.

All the regular solutions for $P$ are either of this form, or can be obtained
via a limiting procedure from $P_a$. It is instructive to look at special cases and present some examples.
The most noted solution, and the only one that can be written in closed form,
is the linear-dilaton background of~\cite{MN}:
\beqs
P&=&\hat{P}\,\equiv\,2N_c\,\r\,.
\label{Eq:MN}
\eeqs
Besides being known analytically, it also has the striking property that any other solution $P$ obeys 
$P>\hat{P}$ for every possible $\r>0$.
In this solution the dilaton $\Phi$ grows indefinitely in the UV. 

A second interesting case is when $c_0=0$ and $c_+$ is positive, so that the
solution is always dominated by the exponential growth in the UV. By inspection, it turns out that
$P_a$ is not an accurate approximation very close to the IR end-of-space, where
the expansion is rather (see~\cite{HNP,GMNP} for details):
\beqs
P_{\ell}&=&h_1\r\,+\,\frac{4h_1}{15}\left(1-\frac{4N_c^2}{h_1^2}\right)\r^3\,+\,\frac{16 h_1}{525}\left(1-\frac{4N_c^2}{3h_1^2}-\frac{32 N_c^4}{3 h_1^4}\right)\r^5\,+\,{\cal O}(\r^7)\,,
\label{Eq:special}
\eeqs
with the constant $h_1\geq 2N_c$. This IR expansion holds for all solutions in which $c_0=0$. Unfortunately, the relation between $h_1$ and $c_+$ is not known in closed form. Lastly, notice that the value $h_1=2 N_c$ reproduces the Maldacena--Nu\~nez background given by $\hat{P}$ above.

On the other hand, if $c_0\neq 0$ we can write the IR expansion as
 \beqs
 P_c&=&c_0\,+k_3 c_0 \r^3+\frac{4}{5}k_3 c_0 \r^5-k_3^2c_0\r^6+\frac{16(2c_0^2 k_3-5k_3 N_c^2)}{105c_0}\r^7\,+\,{\cal O}(\r^8)\,,
 \label{Eq:IRexpansion}
\eeqs
where now $c_0$ and $k_3$ are the free parameters.
Again, the relation between $k_3$ and $c_+$ is not known analytically.
Solutions in this class are more general. Notice  that this parametrization has to be used 
with caution: 
for small enough values of $k_3$, one expects that at some value of $\r$ the solution will become smaller that
$\hat{P}$, yielding a bad singularity. Hence there exists a minimum allowed value of $k_3$,
which depends on $c_0$.

Finally, it is useful to write explicitly the UV expansion. As we said, 
the solution for $P$ cannot asymptote to a constant. 
If the solution asymptotes to a linear $P$, then the dilaton would diverge also linearly, as in~\cite{MN}.
In the following we  only need the expansion obtained
 in the case where $P$ grows exponentially at arbitrarily large
values of the radial coordinate~\cite{HNP}:
\beqs
P_{UV}&=&3c_+ e^{4\r/3}\,+\,4 \frac{N_c^2}{3 c_+} \left(\r^2 - \r + \frac{13}{16}\right) e^{-4 \r/3}\,-
\,\left({8}c_+\r+\frac{c_-}{192c_+^2}\right)e^{-8\r/3}\,+\,{\cal O}(e^{-4\r})\,,
\label{EQ:UVexpansion}
\eeqs
where now the constant $c_-$ is related to $c_0$, the particular form of the relation not known in closed form.

The generic solution will start in the IR with constant $P$, followed by a region where $P$ is linear in $\r$, eventually succeeded by an exponential growth. One or more of these sectors might not be present, depending on the value of the integration constants in the particular solution.

Let us remind the reader about what is known of these distinct regions and what is happening in the dual field theory.
A background quantity that will turn out to be of great importance in the analysis is the
 following, for which we borrow the notation of~\cite{BCNSY}:
\beqs\label{barvev}
M_1&\equiv&4e^{2h-2g}+a^2-1\,=\,\frac{2Q}{P\coth(2\r)-Q}\,.
\eeqs
For $M_1=0$ the BPS equations (and the background solutions) exhibit the $\mathbb{Z}_2$ symmetry
characteristic of the Klebanov-Strassler system. On the contrary, this quantity is non-trivial along the baryonic branch,
as well as in all the solutions of the wrapped-$D5$ system.

For simplicity, in this approximate analysis (but not in the following sections) we set $a=0$. This means that the discussion in the 
rest of this section does not apply in the region near the end of space. Physically, in the dual theory we will be ignoring the formation of the gaugino condensate. We focus our attention on the metric.

\subsubsection{Region with exponential $P$}

Keeping only the leading-order term of the solution in the far UV, controlled by $c_+$, we find 
\begin{equation}
P\,\simeq \, 3\,c_+\,e^{\frac{4}{3}(\r-\r_{\ast})}\,,\qquad\qquad\qquad \qquad\partial_{\r} P \,\simeq \, 4\,c_+\,e^{\frac{4}{3}(\r-\r_{\ast})}\,,
\end{equation}
where $\rho_{\ast}$ is the value of the radial direction above which this approximation is good.
The metric becomes
\beqs\label{expmetric}
\di s^2 &\simeq&\di x_{1,3}^2\,+\,\frac{3}{4}\,c_+\, e^{\frac{4}{3}(\r-\r_{\ast})}\left(\frac{8}{3}\di \r^2\,+\,e_1^2+e_2^2+e_3^2+e_4^2+\frac{2}{3}e_5^2\right)\,\,.
\eeqs
In this expression we see explicitly the form of the natural metric defining $T^{1,1}$, the base of the conifold. Indeed, by changing radial coordinate according to $e^{\frac43\rho}=r^2$ it is easy to see that Eq.~(\ref{expmetric}) is the direct product of four-dimensional Minkowski space and the conifold itself.
In this case, the physics is dominated in the far UV by the insertion of a 
dimension eight operator in the dual field theory~\cite{dimensions}.
As such, backgrounds with this asymptotic behavior are to be understood as
the gravity duals of UV-incomplete field theories.

Nevertheless, as explained for instance in~\cite{EGNP}, the rotation procedure of~\cite{rotation} allows us to construct explicitly the gravity dual of the partial UV-completion of such field theory.
By rotating, the higher-dimensional operator is replaced by an enlarged gauge group (namely, the fact that $F_5$ and $B_2$ are non-trivial in the rotated case translates into the dual field theory having a two-site quiver rather than a simple gauge group). Let us give some details of this procedure. 

The dilaton is
\beqs\label{expdilaton}
e^{4\Phi}&\simeq & 1\,-\,2e^{-4\r}\,+\,{\cal O}(e^{-8\r})\,,
\eeqs
where we set the integration constant $\Phi_0$ so that $\Phi_{\infty}=0$. Using the formulas in~\cite{EGNP}, one can tune the rotation in such a way as to cancel the dimension eight operator from the UV expansion of the rotated background functions. In this instance, the metric is given approximately by
\beqs
\di s^2 &=&e^{2\r}\di x_{1,3}^2\,+\,\frac{3}{4}\,c_+\, e^{-\frac{4}{3}\r_{\ast}}\,e^{-\frac{2}{3}\r}\left(\frac{8}{3}\di \r^2\,+\,e_1^2+e_2^2+e_3^2+e_4^2+\frac{2}{3}e_5^2\right)\,.
\eeqs
Again, in the radial coordinate $e^{\frac43\rho}=r^2$ one can see explicitly that this metric is of the form of a D3-brane
\begin{equation}
\di s^2\,=\,\hat h^{-1/2}\,\di x_{1,3}^2+\hat h^{1/2}\,\di s_6^2
\end{equation}
where the warp factor is $\hat h\sim r^{-6}$ and the transverse space $\di s_6^2$ is once more the conifold. 
An important point is that since we neglected the function $Q$, we are effectively disregarding 
the effects of the non-trivial $F_3$. Hence we are missing with
this rough approximation an important correction:  
the fact that the metric represents the dual 
of a cascading field theory, as in the Klebanov-Strassler background and the baryonic branch.
Nevertheless, the important piece of information for the purposes of this paper 
is that the internal metric is the one of $T^{1,1}$. Notice also that the metric is not AdS, even if we neglect the contribution of the three-form, and
non-vanishing $F_5$ is generated through the rotation.

\subsubsection{Region with linear $P$.}

In the range where $P$ is linear, the following approximations hold
\begin{equation}
P\,\simeq \, 2 N_c\, \r\,,\qquad\qquad\qquad\qquad\partial_{\r} P \,\simeq \, 2N_c\,.
\end{equation}
In this region the most important effect in the dual field theory is
the baryonic VEV, a dimension two condensate, as can be seen by the fact that $M_1$
is not suppressed.
In this case, the metric takes the form
\beqs
\di s^2 &=&e^{\Phi}\,\left[ \di x_{1,3}^2 \,+\,  {N_c}  \di \r^2\,+\, N_c\,\r \left(\frac{}{} e_1^2+e_2^2\right)
\,+\,\frac{N_c}{4} \left(\frac{}{} e_3^2+e_4^2\right)
\,+\,\frac{N_c}{4} \,e_5^2\right]\,.
\eeqs
Now the $\mathbb{Z}_2$ symmetry of the $T^{1,1}$ is broken, in the sense that the two $S^{2}$ described by $(\theta,\phi)$ and $(\tilde\theta,\tilde\phi)$, having different warp factors, cannot be interchanged. This is a direct consequence of the presence of the baryonic VEV.

\subsubsection{Region with constant $P$.}

Let us focus our attention in the region where $P\simeq c_0\gg 2N_c\r$.
In this case $\Phi$ is approximately constant.
A brutal way of treating this system is to set $N_c=0=Q$ in the equations,
in such a way that $\Phi=0$ is  a solution. 
We can thus take
\begin{equation}
P\,=\,c_0\left(\frac{}{}1+e^{4(\r-\r_{\ast})}\right)^{1/3}\,\simeq \,c_0\,,\qquad\qquad\qquad\qquad\partial_{\r}P\,\simeq\,\frac{4}{3}c_0\,e^{4(\r-\r_{\ast})}\,,
\end{equation}
with $\r_{\ast}$ the scale at which the exponential behavior of $P$ finally shows up.
All the forms are trivial, and the only non-trivial background function is the metric. In this limit it can be approximated by:
\beqs
\di s^2 &=&\di x_{1,3}^2\,+\,\frac{c_0}{4} \left(\frac{}{}e_1^2+e_2^2+e_3^2+e_4^2\right)\,
+\,\frac{2}{3}c_0\,e^{4(\r-\r_{\ast})}\left(\frac{}{}\di \r^2+\frac{1}{4}e_5^2\right)\,.
\eeqs
One can see that now the four-dimensional space
described by $\theta$, $\tilde{\theta}$, $\phi$ and $\tilde{\phi}$  is blowing up towards the IR.
This kind of backgrounds have been discussed in various contexts, besides
the one of direct interest for this paper (see for instance~\cite{PZ,CDF}).

The field theory analysis suggests that the dynamics in this region is dominated by the presence of a large 
condensate for an operator of dimension six.
The fact we want to highlight is that in this region
once again the $\mathbb{Z}_2$ symmetry is preserved,
since the effect of $M_1$ is  suppressed.

\section{Probing the geometry: general results.}
\label{Sec:extended}

In this section we present the general formalism with which we will study the brane probes. 
The basic setup we want to investigate is the following.
Suppose one has an extended object (a string or a brane) that is treated as  a probe and assume that there are only two coordinates for which the embedding is determined dynamically. One of these will be the radial coordinate $\r$ and let us refer to the other as $x$. We can parameterize the one-dimensional profile of the probe in the $(\r,x)$ plane in terms of a single variable $\sigma$ as  $x=x(\sigma)$ and $\r=\r(\sigma)$. The class of actions that we will consider reads
\beqs
{\cal S}&=&\frac{T}{2\pi\alpha^{\prime}} \int \di \sigma \sqrt{F^2x^{\prime\,2}+G^2 \rho^{\prime\,2}}\,,
\label{Eq:probe}
\eeqs
where the prime refers to derivatives with respect to $\sigma$ and $T$ is some constant. The functions $F$ and $G$ depend  in general on the radial coordinate but not explicitly on $x$. In particular, this means that $x^{\prime}=0$ is a solution of its own equation of motion, derived from this action.

The form of the action (\ref{Eq:probe}) is the strongest assumption we are going to make. For the system to be described by it, one may need for instance the Wess--Zumino term to vanish, and/or some other embedding coordinate to be fixed dynamically. Both criteria are met in the class of embeddings we are interested in. Once we have an action that reduces to Eq.~(\ref{Eq:probe}), all the analysis we are going to present applies, independently of the nature of the probe and the background it explores.  

From the classical equations of motion one would like to find solutions for which the probe 
has a $U$-shaped form in the $(\r,x)$ plane, reaching out at $\r\rightarrow +\infty$.
The profile is then dictated by the minimum value $\hat{\r}_o$ reached by the probe in the interior of the geometry.

In order to solve the equations, we make use of parameterisation invariance to 
set $\sigma=\rho$, noticing that there must be two branches. Let us define the following quantity:
\begin{equation}\label{Veff}
V_{\rm eff}^2(\r,\hat{\r}_o)\,\equiv\,\frac{F^2(\r)}{F^2(\hat{\r}_o)\,G^2(\r)}\,\left(F^2(\r)-F^2(\hat{\r}_o)\right)\,,
\end{equation}
in terms of which the separation along the $x$ direction between the end-points of the probe at $\r=+\infty$ and its explicit shape are given respectively by
\beqs
L(\hat{\r}_o)&=&2\int_{\hat{\r}_o}^{\r_{\rm U}}\di \tilde{\r}\,\,\frac{1}{V_{\rm eff}(\tilde{\r},\hat{\r}_o)}\,,\label{Lsol}\\[4mm]
x(\r,\hat{\r}_o)&=&\left\{\begin{array}{cc}
\frac{L}{2}-\int_{\hat{\r}_o}^{\r}\di \tilde{\r}\,\,\frac{1}{V_{\rm eff}(\tilde{\r},\hat{\r}_o)}\,,\,&\qquad(x<\frac{L}{2})\cr\\
\frac{L}{2}+\int_{\hat{\r}_o}^{\r}\di \tilde{\r}\,\,\frac{1}{V_{\rm eff}(\tilde{\r},\hat{\r}_o)}\,,\,&\qquad(x>\frac{L}{2})\cr\label{xsol}
\end{array}\right.
\eeqs
The definition of the effective potential (\ref{Veff}) is motivated by the form of the equations of motion, which reduce to $\partial  \rho/ \partial x =\pm V_{eff}$, as shown for instance in \cite{NPR}. Furthermore, the total energy  of the configuration is
\begin{equation}
E(\hat{\r}_o)\,=\,2\int_{\hat{\r}_o}^{\r_{\rm U}}\di \tilde{\r}\,\,\sqrt{\frac{F^2(\tilde{\r})\,G^2(\tilde{\r})}{F^2(\tilde{\r})-F^2(\hat{\r}_o)}}\,,
\end{equation}
obtained by replacing the classical solutions, with the ansatz $\rho=\sigma$, into the action. 
Notice that in general there is a UV divergence, so we used a fixed regulator $\r_{\rm U}$. The physical results are expected to be independent of $\r_{\rm U}$, thus we
may be required to subtract a divergence,
since it is understood that eventually one has to take the limit $\r_{\rm U}\rightarrow +\infty$.

The function $x$ is the actual solution to the classical equations derived from $\cal S$. It can be a complicated function of $\r$ and $\hat{\r}_0$, depending on the background one considers.
In particular, it is possible that $L(\hat{\r}_o)$ turns out not to be monotonic.
If this is the case, then one finds a peculiar situation: there will be several solutions,
characterized by different values of $\hat{\r}_o$, for which $L$ is the same, but in general $E$ is not. This means that the energy, as a function of the separation, would be multivalued. 

Here enters the crucial point we will make use of in the following.
In the spirit of holography, the field theory data is entirely encoded in the boundary values of the relevant functions probing the bulk.
We can think of them as control parameters.
For instance, the separation $L$ between the endpoints in the far UV
is the field theory control parameter for the problem at hand.
Once the UV boundary conditions are specified, the whole configuration is then 
determined by solving the appropriate equations in the bulk. However, in some situations various bulk configurations can satisfy the same UV boundary conditions (control parameters). In this case one must evaluate the action on the various 
classical solutions with the same $L$, and retain only the minimal action one.
The other solutions may be realized as metastable or unstable configurations.
From now on, we refer to the minimal action configuration as {\it stable}
and to the others (if any exists) as {\it unstable}.

A closely related question is that of classical perturbative stability, that is, the absence of tachyons in the spectrum of small fluctuations. Whilst it is clear that non-minimal action configurations are energetically disfavored, this does not preclude them to be physically realized as a metastable state. Conversely, being the embedding with minimal action does not ensure that the spectrum of fluctuations is entirely healthy. 

A radial dependent quantity that plays a prominent role in determining perturbative stability of the embedding is 
\beqs\label{Zsol}
{\cal Z}(\r)&\equiv& \partial_{\r}\left(\frac{G(\r)}{\partial_{\r}F(\r)}\right)\,.
\eeqs
We will derive this function and its efficacy in diagnosing (in)stabilities in the next section. For the time being, let us summarize a set of necessary conditions that must be satisfied by the functions defined above, in order for the
embedding to exist.

\begin{itemize}

\item The function $F(\r)$ must be monotonically increasing. The reason for this requirement can be easily seen
in the definition of $V_{\rm eff}$, and how it enters in Eqs.~(\ref{Lsol})-(\ref{xsol}). If $F(\r)$ is not monotonically increasing,
there will exist values of $\hat{\r}_o$ such that $V_{\rm eff}^2 <0$. 
In particular, this might happen near the end of space, in the presence of a singular behavior
of the background geometry. In this instance, there are no classical solutions
for the probes which reach the end of space, but rather all possible profiles extend only down
to a $\r_{\rm min}$, defined so that $F(\r)$ is monotonically increasing for $\r>\r_{\rm min}$.

\item The effective potential $V_{\rm eff}$ must be such that $V_{\rm eff}\rightarrow +\infty$ 
when $\r\rightarrow +\infty$.
The logic behind this condition is that one aspires to interpret the way the probe joins at infinity in terms of a field theory quantity, and hence one wants the separation $L$ to converge 
when $\r_{\rm U}\rightarrow +\infty$.
In this way $L$ can be thought of as a control parameter in the dual field theory.

\item  
Classically stable solutions must have $\di L/\di \hat{\r}_o\le0$. This is automatically true if
${\cal Z}\le0$. If on the contrary there is a range in which ${\cal Z}>0$, there may be solutions that are perturbativley unstable, 
provided $\hat{\r}_o$ falls in that region.

\end{itemize}

The advantage of considering ${\cal Z}$ should be clear: it is comparatively simple to compute, and if it becomes positive for some value of the radial coordinate, the embedding turns out to be unstable in the vicinity of that region. Conversely, if $\mathcal{Z}$ is negative semidefinite, the embedding of the probe with action (\ref{Eq:probe}) is stable. This is the criterion we will apply in the following.

\subsection{Derivation of ${\cal Z}$.}

Let us show the origin of the function ${\cal Z}$ and its relation with instabilities.
Some of the arguments discussed here can be found also in~\cite{BS,NPR}.
The starting point is the expression we wrote for $L$, for which we assume that $L(\hat{\r}_o)$
is invertible, at least locally. We begin by rewriting
\beqs
L&=&2\int_{\hat{\r}_o}^{\r_{\rm U}}\,\di \r\, \,\frac{G(\r)}{F(\r)}\,\,{\cal K} \left[\frac{F(\r)}{F(\hat{\r}_o)}\right]\,
\eeqs
where the functional ${\cal K}$ is defined as
\beqs
{\cal K}[x]&\equiv& \frac{1}{\sqrt{x^2-1}}\,.
\eeqs
Given that $\r>\hat{\r}_o$ and $F$ is monotonically increasing, this is real and positive definite. After some algebra, which involves integrations by parts where boundary terms must be retained, 
one can write the derivative of the separation as
\beqs\label{Sepder}
\frac{\di L}{\di \hat{\r}_o} &=&2 \lim_{\r_{\rm U}\rightarrow +\infty}
\frac{\partial_{\r}F(\hat{\r}_o)}{F(\hat{\r}_o)}\,
\left\{-\frac{G(\r_{\rm U})}{\partial_{\r}F(\r_{\rm U})}\,{\cal K}\left[\frac{F(\r_{\rm U})}{F(\hat{\r}_o)}\right] 
\,+\,\int_{\hat{\r}_o}^{\r_{\rm U}}\di \r\,{\cal K} \left[\frac{F(\r)}{F(\hat{\r}_o)}\right]
\partial_{\r} \left(\frac{G(\r)}{\partial_{\r}F(\r)}\right)\right\}\,.
\eeqs
It can be seen by a change of variable $\r\rightarrow \log F$ that
convergence at the upper limit of the integral in Eq.~(\ref{Lsol}) together with
divergence of $F$ implies that $(V_{\rm eff}\partial_\r \log (F))^{-1}\rightarrow 0$
as $\r\rightarrow +\infty$. Under this condition (which is satisfied in all cases
relevant to gauge/gravity dualities) a rewriting of the first term in the right-hand side
of Eq.~(\ref{Sepder}) reveals that it must vanish (see also \cite{Wilsonstability}). In this way we obtain
\begin{equation}\label{dLdrho}
\frac{\di L}{\di \hat{\r}_o} \,=\,2 \lim_{\r_{\rm U}\rightarrow +\infty}\frac{\partial_{\r}F(\hat{\r}_o)}{F(\hat{\r}_o)}\,\int_{\hat{\r}_o}^{\r_{\rm U}}\di \r\,\,{\cal K} \left[\frac{F(\r)}{F(\hat{\r}_o)}\right]
\mathcal{Z}\left(\rho\right)\,.
\end{equation}
Notice that $F>0$ by definition, and we already required it to be monotonically increasing, so $\partial_{\r}F>0$. As we said $\mathcal{K}>0$ and thus we conclude that the sign of $\di L/\di \hat{\r}_o$ is governed by $\mathcal{Z}$. In particular, in order for $\di L/\di \hat{\r}_o\le0$, a sufficient condition is $\mathcal{Z}\le0$. On the contrary, if $\mathcal{Z}$ becomes positive in certain range, $\di L/\di \hat{\r}_o$ 
can vanish or become positive for some values of $\hat{\rho}_o$.  
 
A similar exercise for $E$ yields the exact relation
\beqs\label{volume}
\frac{\di E}{\di \hat{\r}_o} &=&F(\hat{\r}_o)\,\frac{\di L}{\di \hat{\r}_o} \,.
\eeqs
To understand the stability conditions of the system it is helpful to consider a thermodynamical analogy. One can identify the function $E$ with the Gibbs free energy $\mathcal{G}(p,T)$, whose natural variable (the pressure $p$, since we work at constant, vanishing temperature) corresponds to the control parameter $L$. In this way, equation (\ref{volume}) is simply the statement
\begin{equation}\label{concavity1}
\frac{\di \mathcal{G}}{\di p} \,=\,V\equiv F(\hat{\r}_o)\ge0\,,
\end{equation}
that we recognize is positive definite as expected. Continuing with the simile, we know that the system will realize the configuration with minimal free energy as a function of the volume at fixed pressure, meaning that we have to single out the solution that minimizes $E(\hat{\r}_o)$. 
In addition, it is also well known that stability requires for $\mathcal{G}$ the concavity condition
\begin{equation}\label{concavity2}
\frac{\di^2 \mathcal{G}}{\di p^2}\le0\qquad\qquad\Leftrightarrow\qquad\qquad \frac{\di V}{\di p}\le0
\end{equation}
that, using our dictionary, can be translated into $\di L/\di \hat{\r}_o\le0$. As we already mentioned, this is verified if $\mathcal{Z}\le0$. Together, the concavity relations (\ref{concavity1}) and (\ref{concavity2}) are the requirements needed for local stability and agree with two of the conditions listed in the previous section. These coincide with the concavity conditions discussed in~\cite{Wilsonstability} for the quark-antiquark potential, dual to a string probe whose action falls in the class (\ref{Eq:probe}). 

In summary, we have presented strong evidence that $\mathcal{Z}\le0$ is a sufficient condition for stability of probe embeddings described by an action of the form ({\ref{Eq:probe}}). Furthermore, in the examples we considered in detail, it turns out to be also necessary. Per contra, for more general embeddings it is likely not to be sufficient. For instance, if the background functions $F$ and $G$ depend on some of the internal angles, the embedding can have instabilities along those directions, as in several examples discussed in~\cite{Wilsonstability}. The diagnostic tool $\mathcal{Z}$ would fail in detecting those symptoms.

\section{(In)stability and chiral-symmetry restoration.}
\label{Sec:applications}

In this section, we apply the criterion we just developed to a special choice of $D7$ embedding 
in the class of backgrounds discussed earlier in the paper.
We will perform the calculations using the numerical solutions for the background functions,
but without approximations. Nevertheless, when useful we will refer to the approximate analysis carried on earlier, 
in order to explain our results. 

\subsection{The $D7$ embedding in the wrapped-$D5$ system.}

We start from the Type IIB backgrounds defined within the wrapped-$D5$ system and examine the embedding of $D7$ branes in the probe approximation.
We adopt an ansatz~\cite{A,KuSo,DKS} according to which the $D7$ fills the four Minkowski coordinates plus the radial direction, but 
also an internal three-dimensional manifold, which we choose to be given by the coordinates
$\tilde{\theta}$, $\tilde{\phi}$ and $\psi$. 

The transverse space is spanned by the remaining two-sphere coordinates $\theta$ and $\phi$. As shown in~\cite{KuSo}, it is consistent to assume that their profile depends just on the embedding coordinate $\sigma$ and not on the rest of the angles. The only non-trivial form in the system is $F_3$, so the action for the $D7$ branes reduces to the DBI part and we have to solve the equations for the profile of $\r(\sigma)$, $\phi(\sigma)$ 
and $\theta(\sigma)$. The DBI action can be computed from the ten-dimensional string-frame metric for the wrapped-$D5$ system with $\alpha^{\prime} g_s=1$, that reads
 \beqs
\di s^2 &=&e^{\Phi} \Big[dx_{1,3}^2 + 
e^{2k}d\rho^2
+ e^{2 h}
(e_1^2+e_2^2) +\frac{e^{2 {g}}}{4}
\left((e_4+a\,e_2)^2
+ (e_3+a\,e_1)^2\right)
 + \frac{e^{2 k}}{4}
e_5^2\Big]\,, 
\eeqs
For the embedding described above, substituting the determinant of the induced metric into the DBI and performing the integral over the angular variables we arrive to the action~\cite{A}
\beqs
{\cal S}_{D7} &\sim& 
\int \di^4 x\,  \di \sigma  \sqrt{e^{4g+4k+6\Phi}\r^{\prime\,2}+e^{4g+2k+2h+6\Phi}\left(\theta^{\prime\,2}+\sin^2\theta\,\phi^{\prime\,2}\right)}\,,
\eeqs
where prime denotes derivatives with respect to $\sigma$ and we have ommited an irrelevant constant. Notice that the $SO(3)$ symmetry of the sphere described by $\theta$ and $\phi$ is unbroken, hence the problem reduces essentially to find the geodesics on the two-sphere. Among all the possible equivalent solutions it is convenient to study the configuration with $\theta=\frac{\pi}{2}$, as in~\cite{KuSo,A}.

The crucial point is that once we fix a geodesic, the action falls into the class (\ref{Eq:probe}), where $\phi$ takes the role of the arbitrary coordinate $x$, with the subtlety that the angle $\phi$ is bounded by construction (we will denote $\bar{\phi}$ the value of the angular
separation between the end-points of the embedding, which corresponds to $L$ in the general discussion).
 As a consequence, all the results of the previous section automatically apply, the background functions being
\begin{equation}
F^2\,=\,e^{4g+2k+2h+6\Phi}\,,\qquad\qquad\qquad\qquad G^2\,=\,e^{4g+4k+6\Phi}\,.
\end{equation}
At this point, we can replace the expressions for the warp factors, in terms of the
functions $P$ and $Q$:\beqs
F^2&=& \frac{\sqrt{2} \,e^{6 \Phi_0} \sinh
   (2 \r)}{\sqrt{P^{\prime} P^2-P^{\prime} Q^2}}\, \left(Q+P\, \sinh (4 \r)-Q\,\cosh (4 \r)\right)\,,\\[2mm]
G^2&=&
4 \sqrt{2} \,P^{\prime\,2} \left(Q\,\sinh (2 \r)-P \cosh (2 \r)\right)^2
  \frac{e^{6 \Phi_0} \sinh (2 \r)}{ \left(P^{\prime}
   P^2-P^{\prime} Q^2\right)^{3/2}}\,.
\eeqs
An important quantity is the asymptotic value of the function $F$ in the IR, as this can be thought of as an effective tension for the brane. Using the expansion (\ref{Eq:special}), in the general wrapped-D5 case we find
\beq
F_{IR}=\frac{2\times 2^{3/4}\, e^{3\F_0}}{h_{1}^{1/4}}\,\,\rho\,\,\left(1+\frac{56\,N_c^2-60 \,h_{1}\,N_c+ 66\, h_{1}^2}{45 h_{1}^2}\,\r^{2}\right)+O\left(\r^5\right)\,,
\label{FIRMN}
\eeq
where the Maldacena--Nu\~nez solution can be recovered by setting $h_1=2\,N_c$. On the other hand, for walking backgrounds we use the expansion (\ref{Eq:IRexpansion}), yielding
\beq
F_{IR}=\frac{2\times2^{3/4}}{3^{1/4}}\frac{e^{3\F_0}}{(c_0 \,k_3)^{1/4}}\,\,\rho^{1/2}\,\,\left(1+\frac43\,\r^2 + \frac{3\,c_0\,k_3-8\,N_c}{6 \,c_0}\,\r^3\right)+O\left(\r^{9/2}\right)\,.
\label{FIRMN}
\eeq
In both cases it is clear that, at the end-of-space, $F(0)=0$ and thus the effective tension vanishes. This has decisive repercussions for the type of embeddings we are allowed to consider. 

\subsection{Solutions with linear $P$.}

\begin{figure}
\begin{center}
\begin{picture}(320,170)
\put(-90,0){\includegraphics[height=6cm]{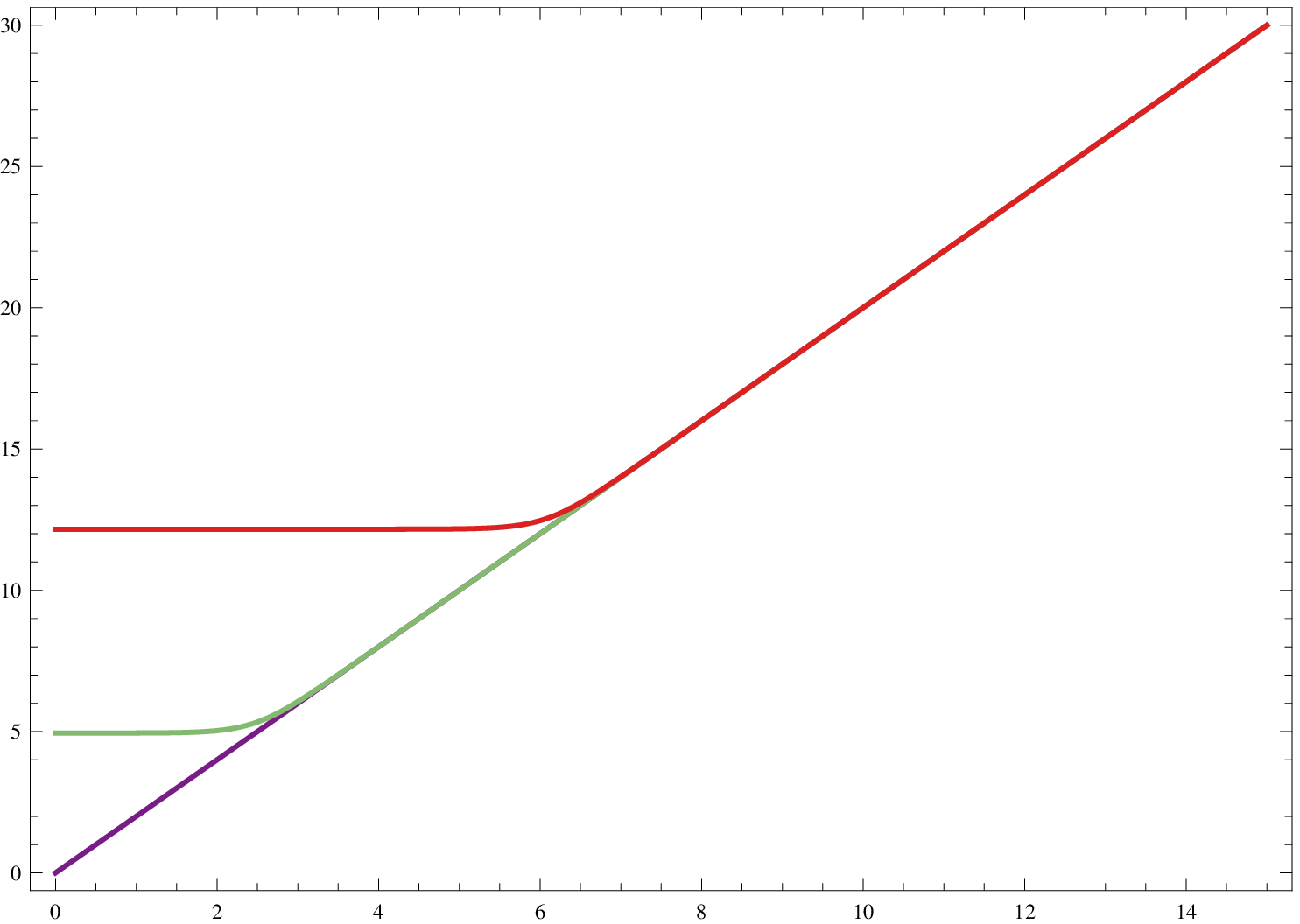}}
\put(-98,165){$P$}
\put(150,-3){$\r$}
\put(-75,16){{\scriptsize{${\hat{P}}$}}}
\put(-75,40){{\scriptsize{$P(\r_{*}\simeq 3)$}}}
\put(-75,76){{\scriptsize{$P(\r_{*}\simeq 6)$}}}
\put(170,0){\includegraphics[height=6cm]{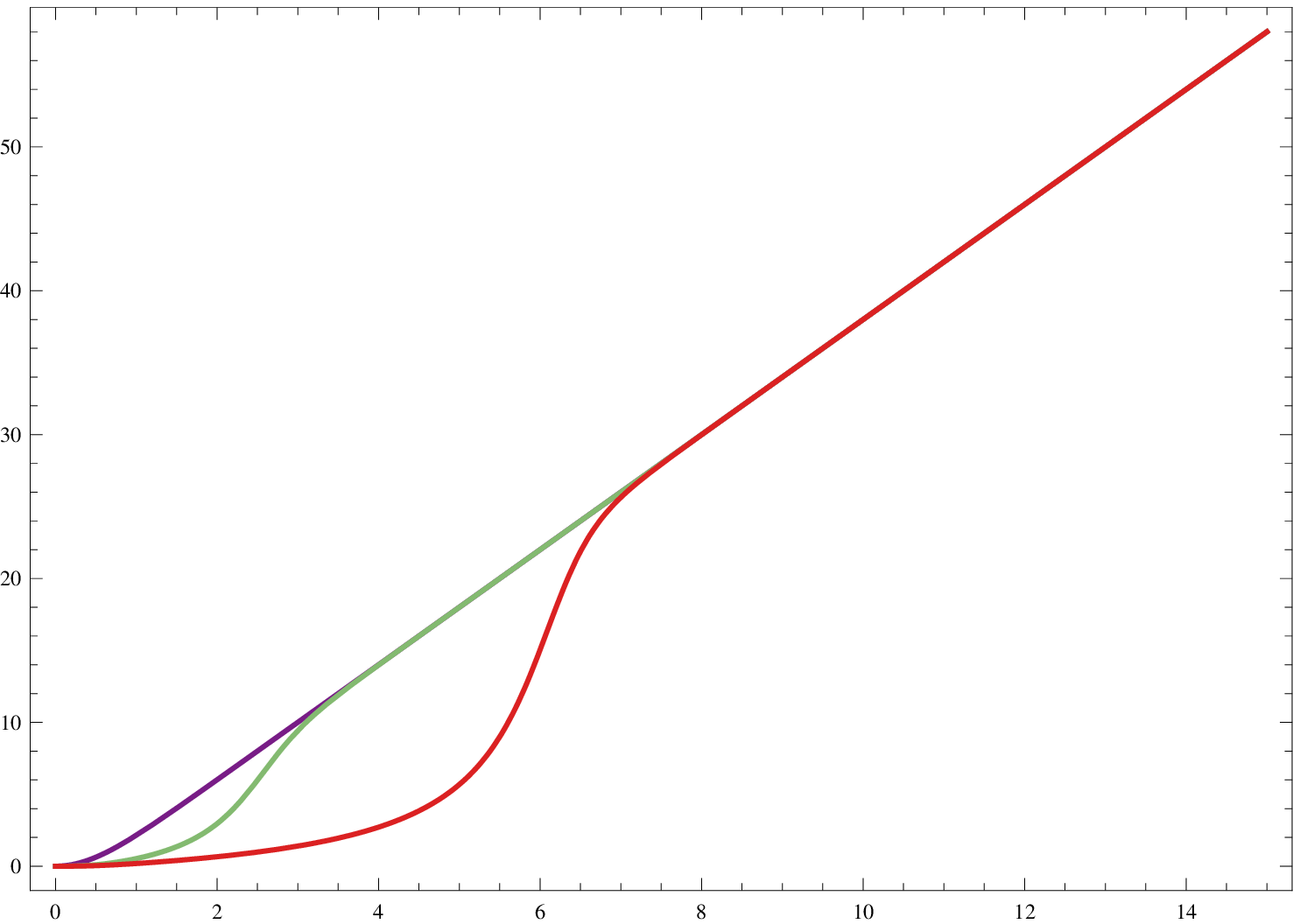}}
\put(160,165){$M_1$}
\put(410,-3){$\r$}
\put(195,20){{\scriptsize{${\hat{P}}$}}}
\put(220,61){{\scriptsize{$P(\r_{*}\simeq 3)$}}}
\put(305,121){{\scriptsize{$P(\r_{*}\simeq 6)$}}}
\end{picture} 
\caption{The functions $P$ (left panel), and  $M_1$ (right panel) 
for three examples of backgrounds in which $P$ is linear for large $\r$. Remember that $\hat{P}$ denotes the Maldacena--Nu\~nez solution.}
\label{Fig:plotMNpM1}
\end{center}
\end{figure}

We focus first on solutions to the wrapped-$D5$ system in which $P$ and the dilaton are linear in the far UV.
We show in Fig.~\ref{Fig:plotMNpM1} three examples of backgrounds of this type.
As we mentioned, this is a one-parameter class of solutions to the master equation.
The parameter can be chosen to be $\r_{\ast}$, the scale below which $P$ is approximately constant (the walking region) and above which it is roughly linear. The presence of this scale is apparent in the quantities we plot. 
The three curves correspond to the Maldacena--Nu\~nez background $\hat{P}$,
and two solutions with $\r_{\ast}\simeq 3$ and $\r_{\ast}\simeq 6$.
We also show, in the right panel of Fig.~\ref{Fig:plotMNpM1}, the baryonic VEV $M_1$ defined in Eq.~(\ref{barvev}).
Notice that in the walking region $M_1$ is suppressed with respect to the $\hat{P}$ case,
as a consequence of the fact that the presence of the VEV for the six-dimensional operator is partially restoring the 
$\mathbb{Z}_2$ symmetry between the two $S^2$ inside the $T^{1,1}$.

\begin{figure}[h]
\begin{center}
\begin{picture}(320,170)
\put(-100,0){\includegraphics[height=6cm]{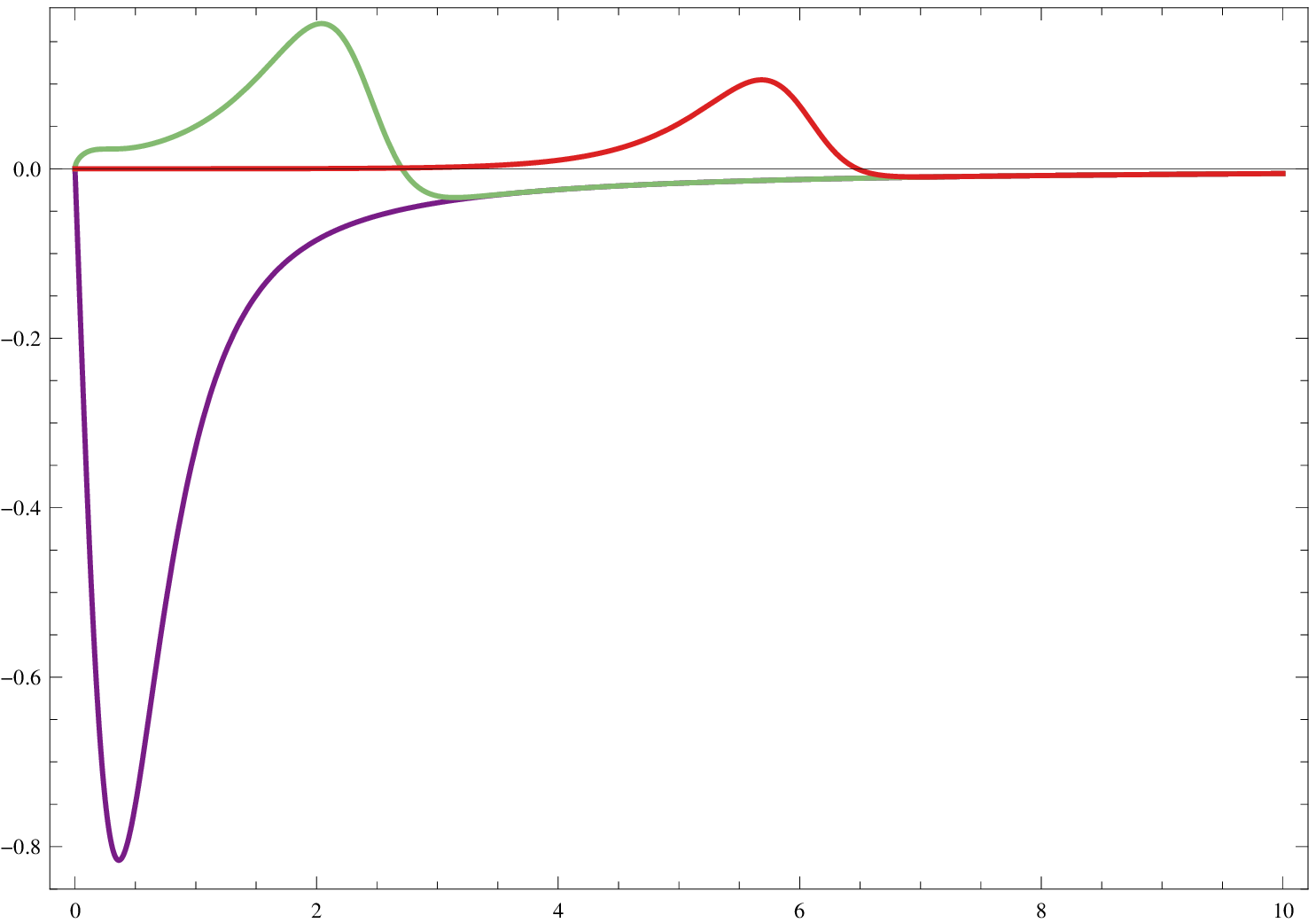}}
\put(-100,162){$\mathcal{Z}$}
\put(143,-3){$\r$}
\put(-60,85){{\scriptsize{${\hat{P}}$}}}
\put(-28,152){{\scriptsize{$P(\r_{*}\simeq 3)$}}}
\put(52,152){{\scriptsize{$P(\r_{*}\simeq 6)$}}}
\put(170,0){\includegraphics[height=6cm]{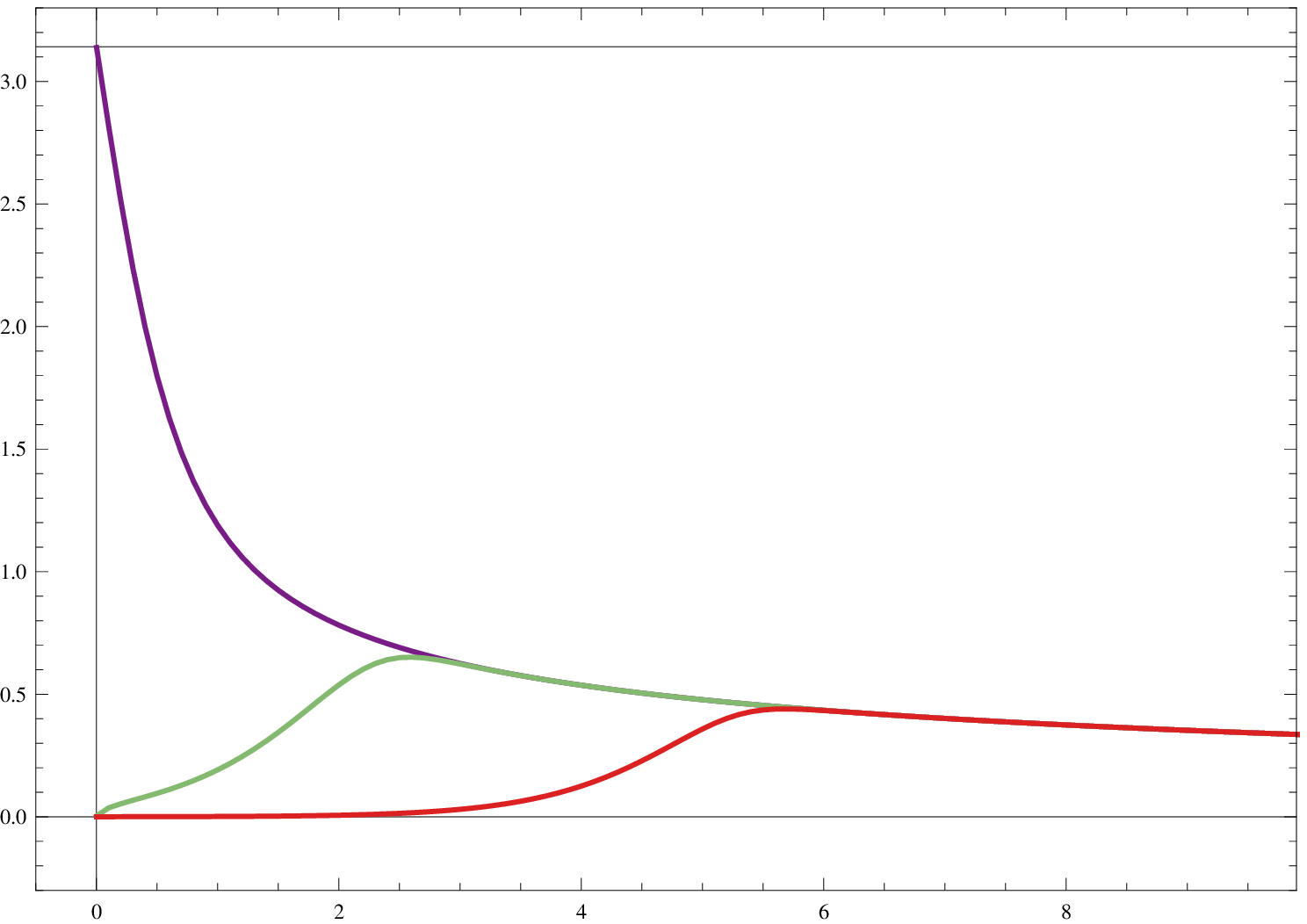}}
\put(167,165){$\bar{\phi}$}
\put(410,-3){$\hat{\r}_0$}
\put(201,100){{\scriptsize{${\hat{P}}$}}}
\put(261,50){{\scriptsize{$P(\r_{*}\simeq 3)$}}}
\put(346,41){{\scriptsize{$P(\r_{*}\simeq 6)$}}}
\end{picture} 
\caption{For the three examples of backgrounds in Fig.~\ref{Fig:plotMNpM1}, 
the function $\mathcal{Z}(\r)$ (left panel) and  the angular separation $\bar{\phi}$ as a function of the 
turning-point $\hat{\r}_o$. In the thermodynamic analogy, this second curve corresponds to the isothermal.}
\label{Fig:plotMNzeL}
\end{center}
\end{figure}

In Fig.~\ref{Fig:plotMNzeL} we display the functions ${\cal Z}(\r)$ (left panel)
and $\bar{\phi}(\hat{\r}_o)$, obtained for the same three examples as in Fig.~\ref{Fig:plotMNpM1}.
For the background given by $\hat{P}$, we find that ${\cal Z}\leq 0$ for every $\r$.
The stability analysis introduced earlier tells us that the embedding we are considering is stable,
in the sense that we do not expect tachyonic excitations to exist.
For every possible choice of the control parameter $\bar{\phi}$ there exists a unique $\hat{\r}_o$ for which the U-shaped embedding satisfies the UV boundary conditions.

In contrast, for the walking solutions the function ${\cal Z}$ becomes positive
for $\r\lsim \r_{\ast}$, and hence we expect those embeddings to be perturbatively unstable.
Indeed, the function $\bar{\phi}(\hat{\r}_o)$ is not invertible: there exist two different choices of 
$\hat{\r}_o$ corresponding to the same value of the control parameter $\bar{\phi}$,
which means that only the one with minimum energy is the classical configuration. 
Furthermore, for a finite value of $\r_{\ast}$ there is a finite maximum value $\bar{\phi}_m(\r_{\ast})<\pi$
of the control parameter above which the U-shaped embedding does not exist.

\begin{figure}[h]
\begin{center}
\begin{picture}(320,130)
\put(-100,3){\includegraphics[height=3.8cm]{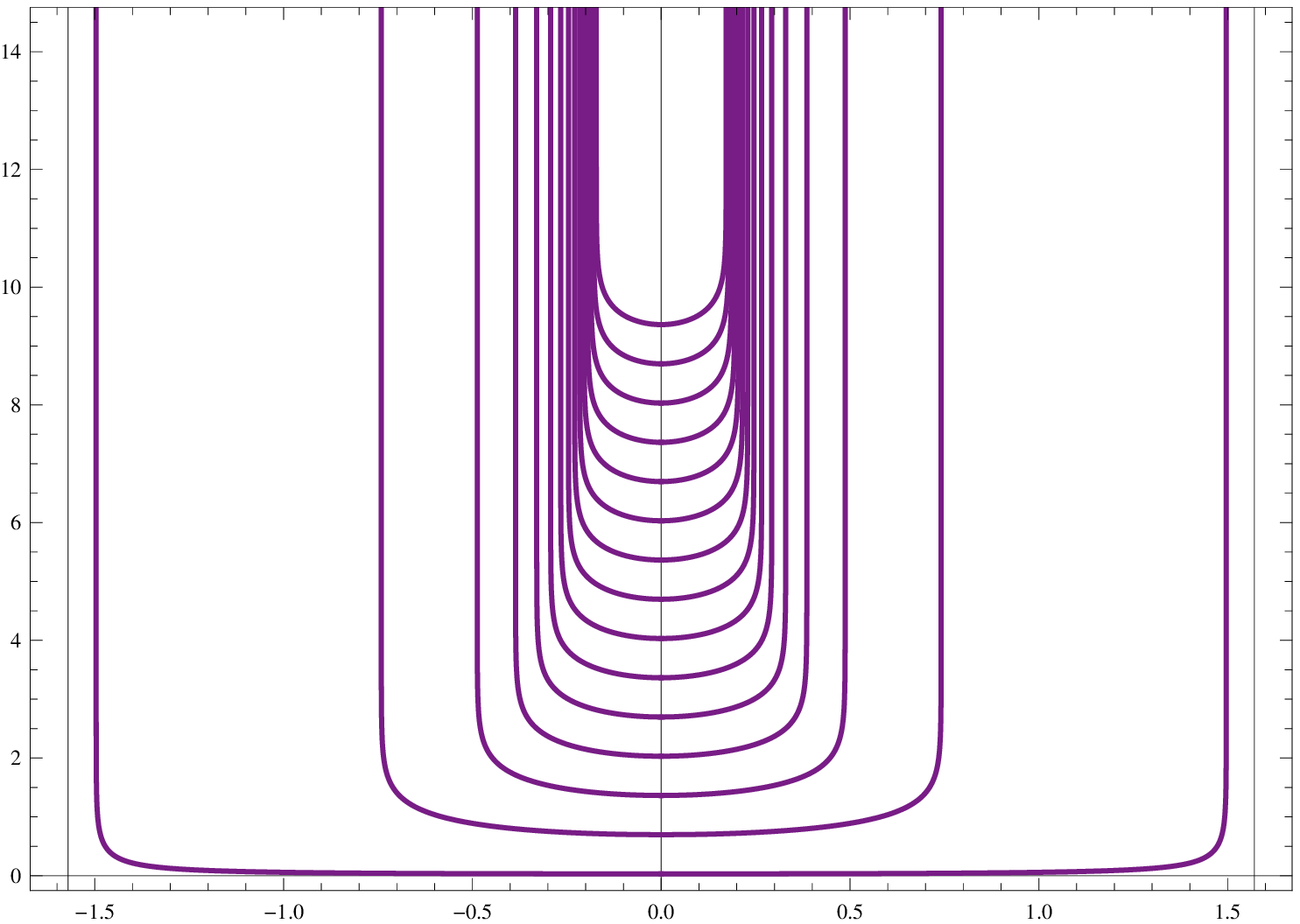}}
\put(-105,100){$\r$}
\put(40,-3){${\phi}$}
\put(80,3){\includegraphics[height=3.8cm]{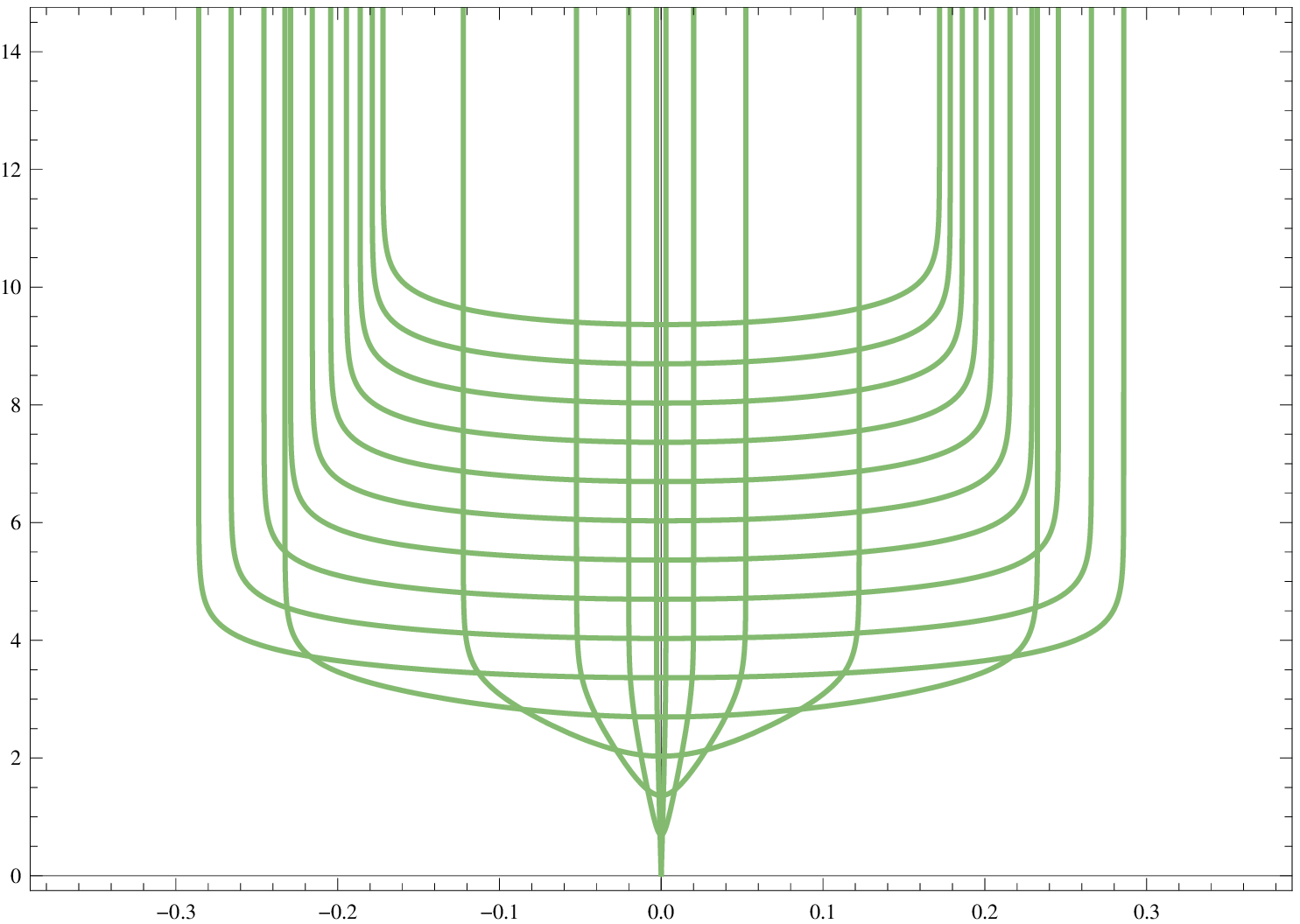}}
\put(75,100){$\r$}
\put(220,-3){${\phi}$}
\put(260,3){\includegraphics[height=3.8cm]{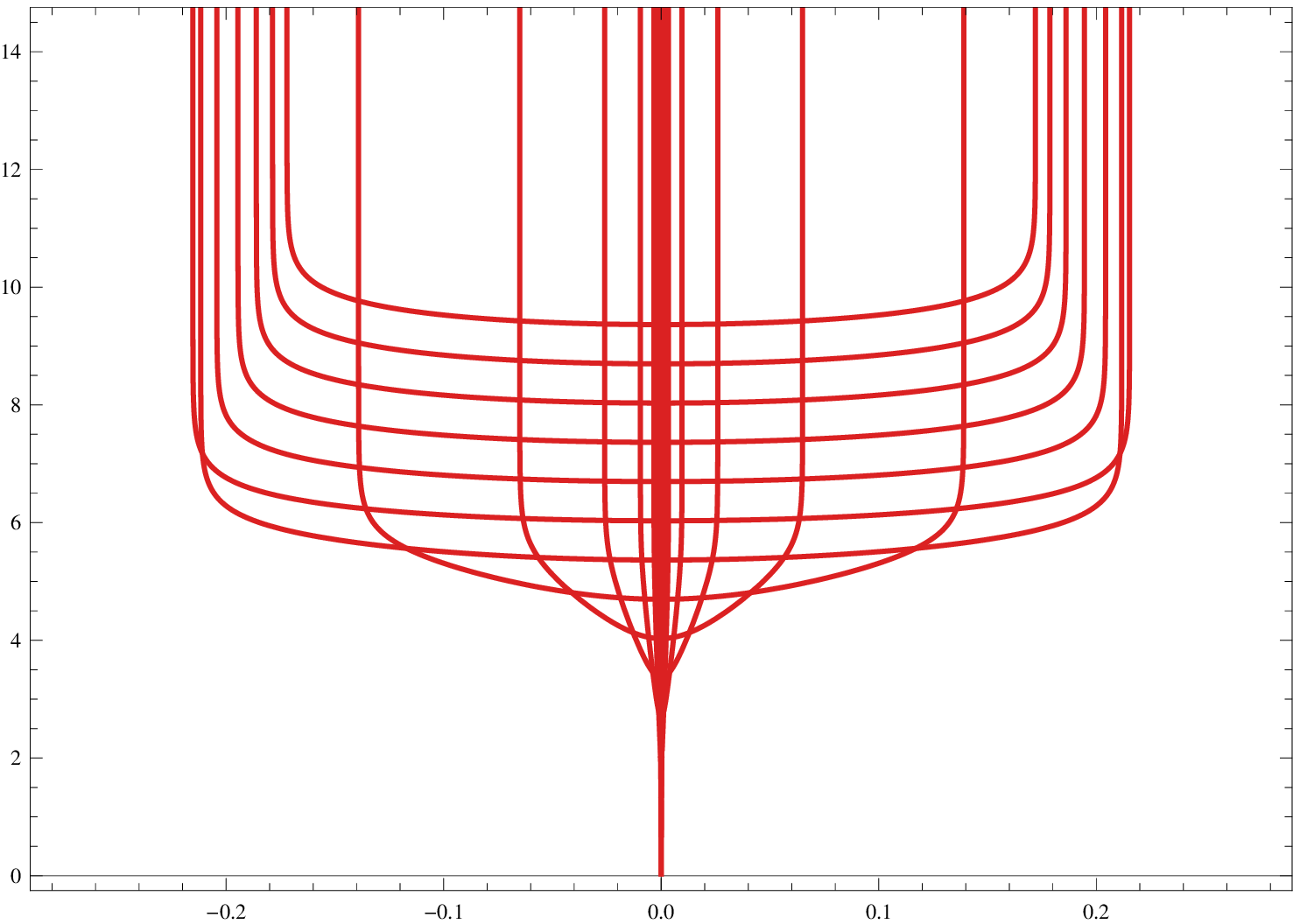}}
\put(255,100){$\r$}
\put(400,-3){${\phi}$}
\end{picture} 
\caption{The $U$-shaped embeddings in the three examples of backgrounds
from Fig.~\ref{Fig:plotMNpM1}, in the plane $({\phi},\r)$, for various choices of $\hat{\r}_o$.
Left to right, we show the results for
the Maldacena--Nu\~nez solution $P=\hat{P}$, for a walking background with $P(\r_{\ast}\simeq 3)$
and for a walking background with $P(\r_{\ast}\simeq 6)$.}
\label{Fig:plotMNEmb}
\end{center}
\end{figure}

The existence of a maximum value of $\bar{\phi}$ opens another problem, i.e., what happens if we choose our control parameter $\bar{\phi}>\bar{\phi}_m(\r_{\ast})$.
In order to answer this question, we must look at the shape of the embedding 
in the $(\r,\phi)$-plane, which we show for a sample of choices of $\hat{\r}_o$ in Fig.~\ref{Fig:plotMNEmb}.
In the Maldacena--Nu\~nez background (left panel of Fig.~\ref{Fig:plotMNEmb}), the embedding is reminiscent of the Sakai--Sugimoto case.
Embeddings that  probe only the UV of the geometry realize small values of $\bar{\phi}$,
while for larger values of $\bar{\phi}$ the turning point of the embedding falls deeper in the IR,
until the antipodal configuration with $\bar{\phi}=\pi$ effectively reaches the end of the space.
There is an important difference with Sakai--Sugimoto though: since the transverse $S^2$ does not shrink
to zero size at the end of the geometry, the antipodal configuration consists 
at $\r=0$ of an arc along the equator of the sphere.
This is a significant fact, which we will  comment about later on.
The essential point is that all of the embeddings in the $\hat{P}$ background are completely smooth and stable.

The situation for walking backgrounds (middle and right panel of Fig.~\ref{Fig:plotMNEmb})
is notably different.
As long as $\hat{\r}_o>{\r}_{\ast}$, the embedding is equivalent to the $\hat{P}$ case.
On the contrary, when we choose $\hat{\r}_o<\r_{\ast}$, the shape of the embedding changes in a
significant way: not only is $\bar{\phi}$ becoming smaller, as seen also in the right panel of 
Fig.~\ref{Fig:plotMNzeL}, but also a non-trivial feature emerges 
at the turning point of the embedding. A similar property was highlighted 
in the study of Wilson loops on the same backgrounds in~\cite{NPR}.
For $\hat{\r}_o\rightarrow 0$ (and contrary to the case of the Wilson loop)
the profile degenerates into a cusp, the angular separation in the UV vanishes
($\bar{\phi}\rightarrow 0$), and effectively the embedding morphs into two straight lines
on top of each other in the $(\r,\phi)$ plane.

In practice, this means that what starts as a U-shaped, connected solution with
finite angular separation $\bar{\phi}$ continuously degenerates, for $\hat{\r}_o\rightarrow 0$, into 
a disconnected configuration in which there are two, independent embeddings.
This allows us to compare directly the energy of the connected configurations
with disconnected ones, by which we mean an embedding in which two 
independent $D7$-branes wrap the same internal and external portions of the space and extend along the whole radial direction at fixed angles $\theta$ and $\phi$. This class of solutions to the equation of motion is sensible due to the vanishing of the tension $F(\hat{\r}_o)$ at the end of space, signaling that the compact space that the branes wrap effectively collapses.
We stress that the comparison between the energy of one U-shaped configuration and 
two disconnected ones is made possible by the fact that the former degenerates into a special case of the
latter (in which $\phi$ is the same for the two branches), which allows to fix an otherwise undetermined overall 
additive constant. We also emphasize that the legitimacy of this procedure rests ultimately on the
fact that the brane becomes tensionless at the end of space~\footnote{
See the critical discussions in~\cite{Bak:2007fk} in which 
the exchange of bulk supergravity modes between the disconnected objects is considered, 
and the famous results in~\cite{Gross:1998gk} where it
is shown how in particular cases the resulting non-perturbative effects are captured 
by special connected configurations of extended objects. 
These special configurations arise in a number of contexts, including 
for instance the background often referred to as QCD$_3$~\cite{Armoni:2013qda},
and here correspond to connecting the two disconnected configurations
via a branch that is localised at $\r=0$ and extends along the angle $\phi$.}.

\begin{figure}
\begin{center}
\begin{picture}(320,170)
\put(-100,0){\includegraphics[height=6cm]{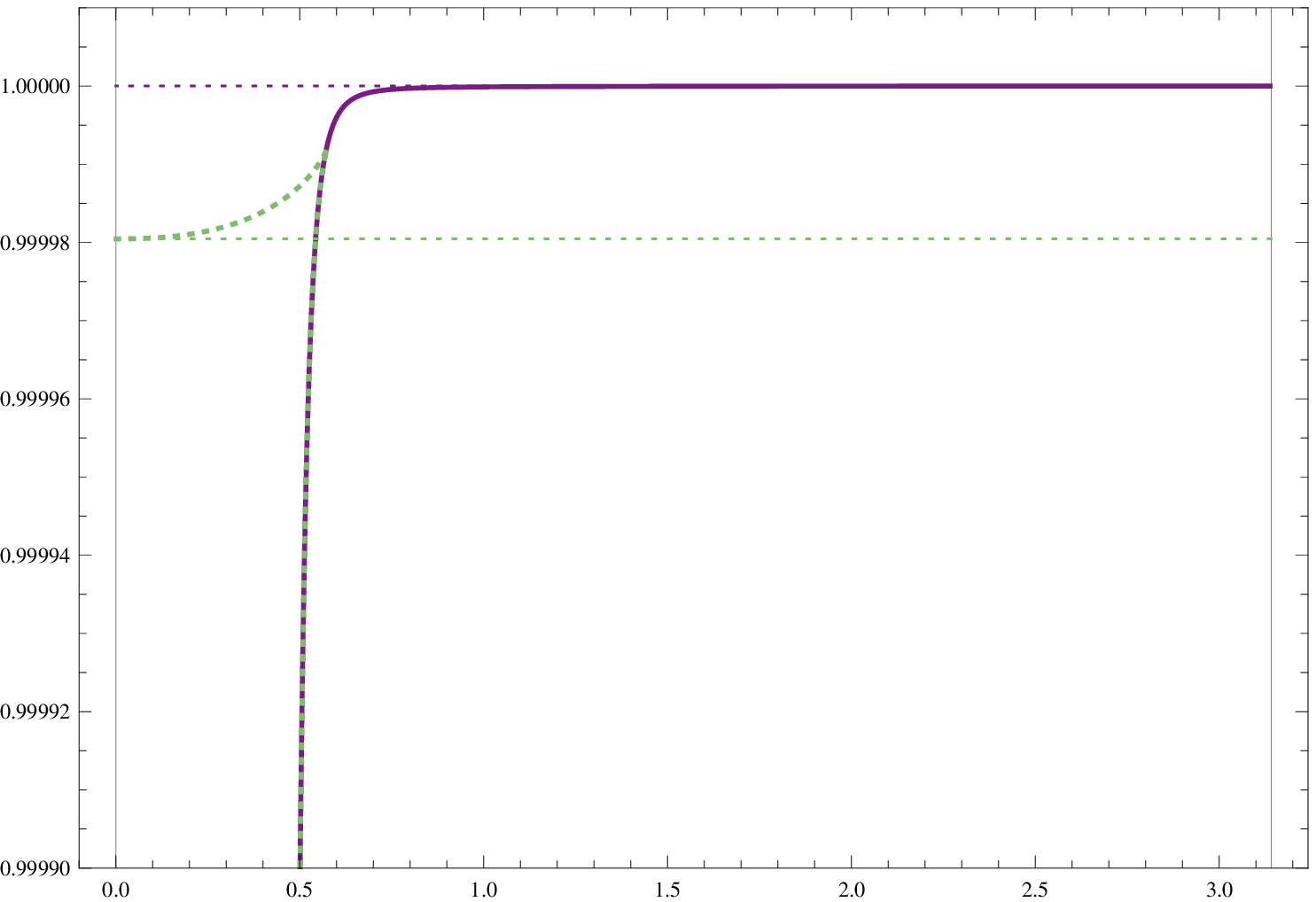}}
\put(-95,165){$E$}
\put(143,-3){$\bar{\phi}$}
\put(-38,116){$\bar{\phi}_c$}
\put(-42,124){$\bullet$}
\put(-34,140){$\bar{\phi}_m$}
\put(-40,141){$\bullet$}
\put(15,15){\includegraphics[height=2.75cm]{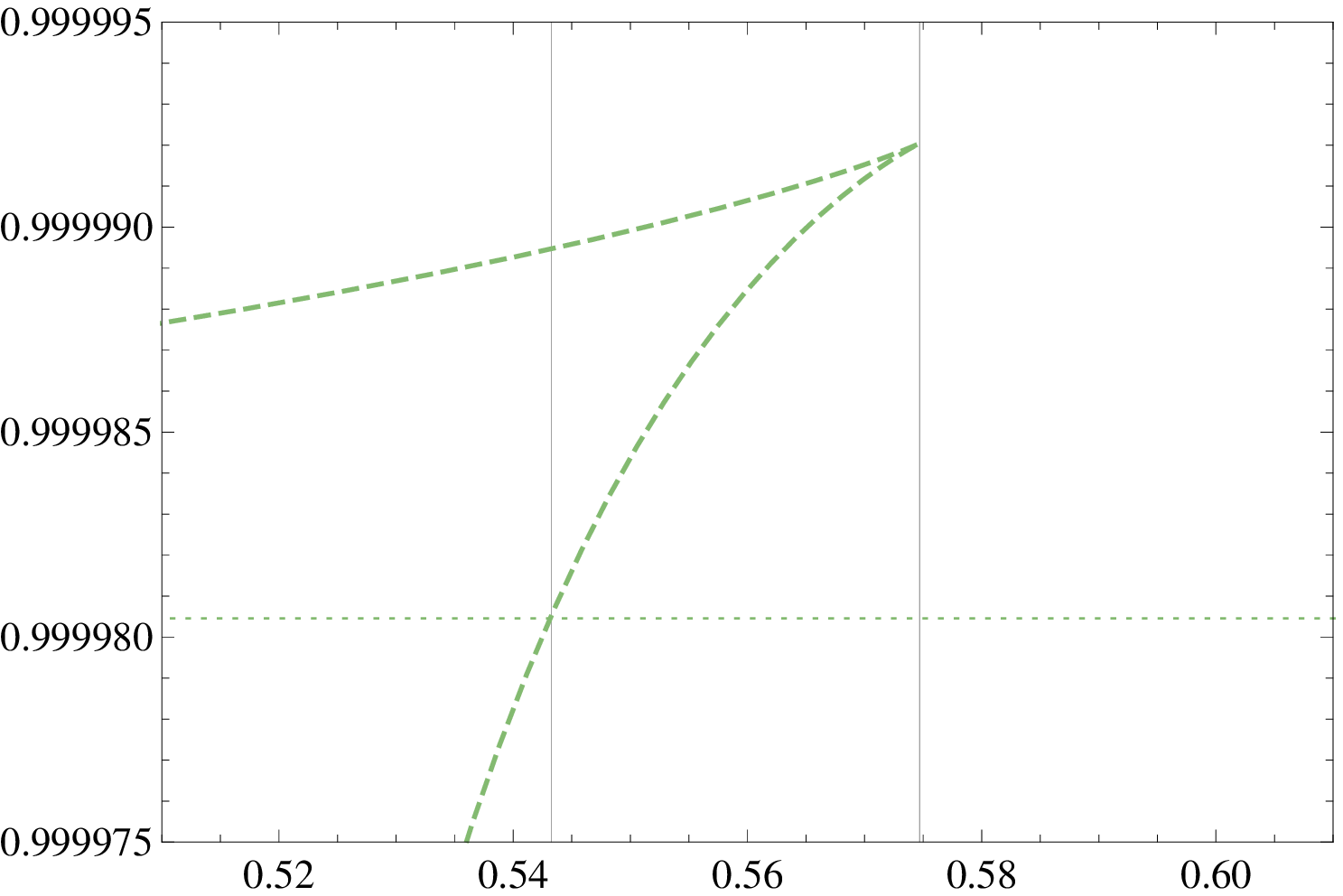}}
\put(160,0){\includegraphics[height=6cm]{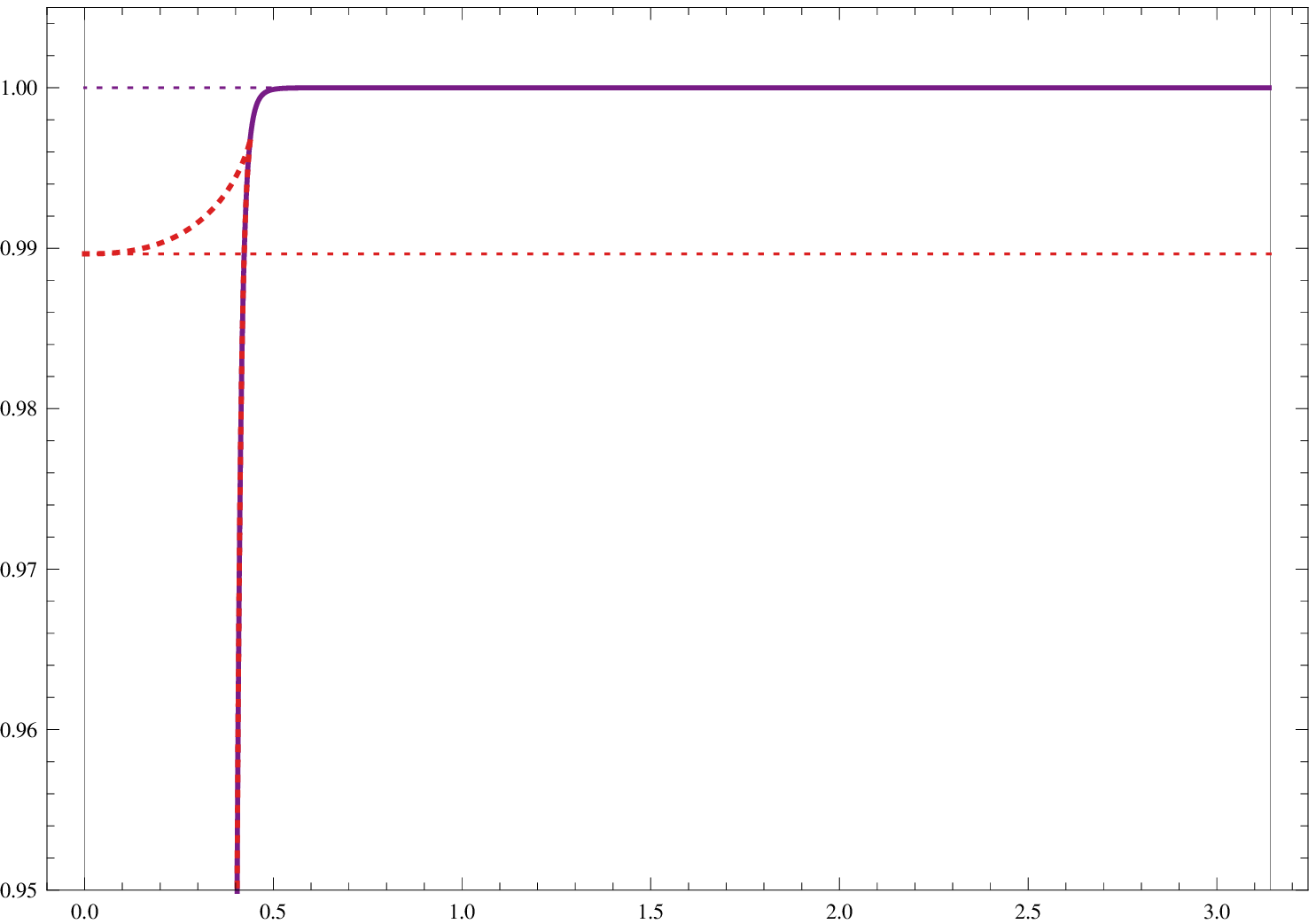}}
\put(160,165){$E$}
\put(403,-3){$\bar{\phi}$}
\put(208,114){$\bar{\phi}_c$}
\put(203,123){$\bullet$}
\put(209,140){$\bar{\phi}_m$}
\put(204,141){$\bullet$}
\end{picture} 
\caption{The energy $E$ as a function of $\bar{\phi}$ for  walking backgrounds with $\r_{\ast}\simeq 3$ (left panel)
and $\r_{\ast} \simeq 6$ (right panel), compared to the MN case.  In the plots, the thin dashed lines represent
 the disconnected solutions and the solid thick line represents the connected ones. The zoom shows a detail of the cusp, where the values of $\bar{\phi}_c$ and $\bar{\phi}_m$ are visible.
  A UV cutoff of $\r=8$ is used when performing the $E$ integral and it has been 
  renormalised such that the disconnected embedding in the MN case has unit energy.}
\label{Fig:plotwalkEL}
\end{center}
\end{figure}

For any given choice of the control parameter the disconnected solution reaching the end of space exists.
Its energy does not depend on $\bar{\phi}$, since at the order we are working the two 
separate branches do not interact.
We can interpret the connected configuration as chiral symmetry breaking,
while the disconnected one corresponds to a chiral symmetry restored phase.
Hence, for walking backgrounds there exists a third possible classical configuration,
and we must ask which one is realised in practice by considering the total energy as displayed in Fig.~\ref{Fig:plotwalkEL}. This is akin to the $\mathcal{G}(p)$ curve in the thermodynamic analogy.

From the figure we see that there is another special value of $\bar{\phi}$,
which we call $\bar{\phi}_c<\bar{\phi}_m$,  at which the curve  $E(\bar{\phi})$ 
representing the disconnected configuration intersects the connected one.
Let us explain what is happening as a function of the control parameter.
For $0\leq \bar{\phi} < \bar{\phi}_c$, there exist three classical solutions.
The minimal energy one is the connected configuration that has a large value of $\hat{\r}_o>\r_{\ast}$.
The other connected configuration, which has a $\hat{\r}_o<\r_{\ast}$ is actually a maximum of the energy,
which explains its tachyonic nature. The disconnected solution is not tachyonic,
yet it happens to have energy larger than the connected one.

For $\bar{\phi}_c < \bar{\phi} < \bar{\phi}_m$, the three classical solutions still exist,
but now the disconnected solution becomes the global minimum.
This is a first-order phase transition taking place at $\bar{\phi}=\bar{\phi}_c$:
as long as we choose a small value of $\bar{\phi}$, the system of probes prefers to realise the
chiral symmetry breaking phase, while for large values of $\bar{\phi}$ the symmetry is restored.
For $\bar{\phi}>\bar{\phi}_m$ only the disconnected configuration exists.
Notice that $E(\bar{\phi})$, constructed by taking the absolute minimum of
the allowed classical configurations, is a continuous function, not differentiable at $\bar{\phi}_c$,
which is the characterisation of a first-order phase transition. This is analogous for instance to the gas/liquid phase transition in the Van der Waals gas as seen in $\mathcal{G}(p)$.

The conclusion is that for all backgrounds with $P$ asymptotically linear there exist stable configurations
of the embedding we are studying, for any choice of $\bar{\phi}$.
However, as a function $\bar{\phi}$, we find a first-order phase transition,
the value $\bar{\phi}_c$ at which it occurs depending on the scale $\r_{\ast}$.

One can think of this phenomenon as the formation of a symmetry-breaking condensate,
 in the presence of an explicit symmetry breaking deformation.
In the Maldacena--Nu\~nez case, the condensate forms for any value of the explicit 
symmetry-breaking term. In the walking backgrounds, the condensate forms only if the source of
explicit symmetry breaking is large enough, that is, if $\bar{\phi}<\bar{\phi}_c$.

Let us add a remark about the geometric properties of this system. It can be shown~\cite{EGNP} that only when $P$ is linear does the manifold wrapped by the $D7$-branes (spanned by $\tilde{\theta}$, $\tilde{\phi}$ and $\psi$) correspond to a round sphere. Any deviation from the linear behavior yields the squashing of the $S^3$. What we are finding is that such deformation has a proclivity for producing instabilities in the U-shapped embedding. Note also that the linear Maldacena--Nu\~nez background  has the largest value of the baryonic VEV $M_1$, correlated with the breaking of $\mathbb{Z}_2$. It appears that restoration of this symmetry ---  as in the walking backgrounds --- also tends to destabilize the configuration.

We close this subsection by highlighting that walking backgrounds with linear $P$ 
are those for which a light scalar glueball has been identified in~\cite{ENP,EP}.
As we have shown, this is also the case in which U-shaped embeddings exist and are stable. It would be very interesting to repeat for these solutions
the construction of a semi-realistic model of technicolor along the lines of~\cite{ASW1},
and compute the $S$ parameter as a function of $\bar{\phi}$.

\subsection{Solutions with exponential $P$.}

We consider now backgrounds in which $P$ is exponential at large $\r$, focusing on a subclass of such solutions: those for which the approximation
$P\gg \hat{P},Q$ holds for all $\r$.
These are the backgrounds originally considered in~\cite{ASW1}, for which the $D7$ embedding 
is known to be classically unstable~\cite{CLV}.
All other classes of solutions within the wrapped-$D5$ system, even after the rotation procedure has
been applied, yield pathologies for this embedding of probe $D7$, as will be shown 
elsewhere~\cite{DS}.
Here we want to explain the origin of the instability found in~\cite{CLV}.
We will make use of the numerical solution for $P$, without any approximation.

\begin{figure}
\begin{center}
\begin{picture}(320,170)
\put(-90,0){\includegraphics[height=6cm]{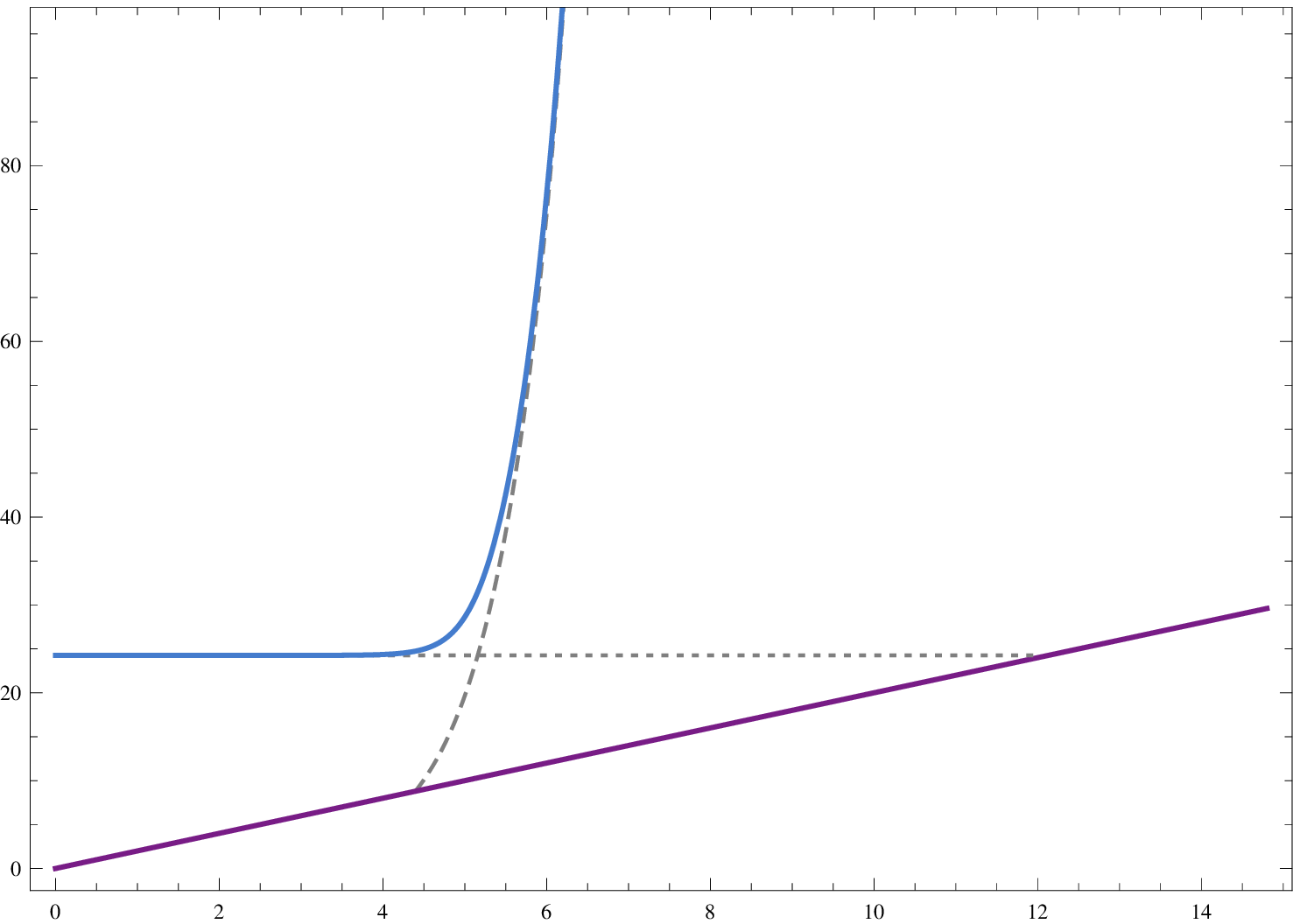}}
\put(-98,165){$P$}
\put(150,-3){$\r$}
\put(170,0){\includegraphics[height=6cm]{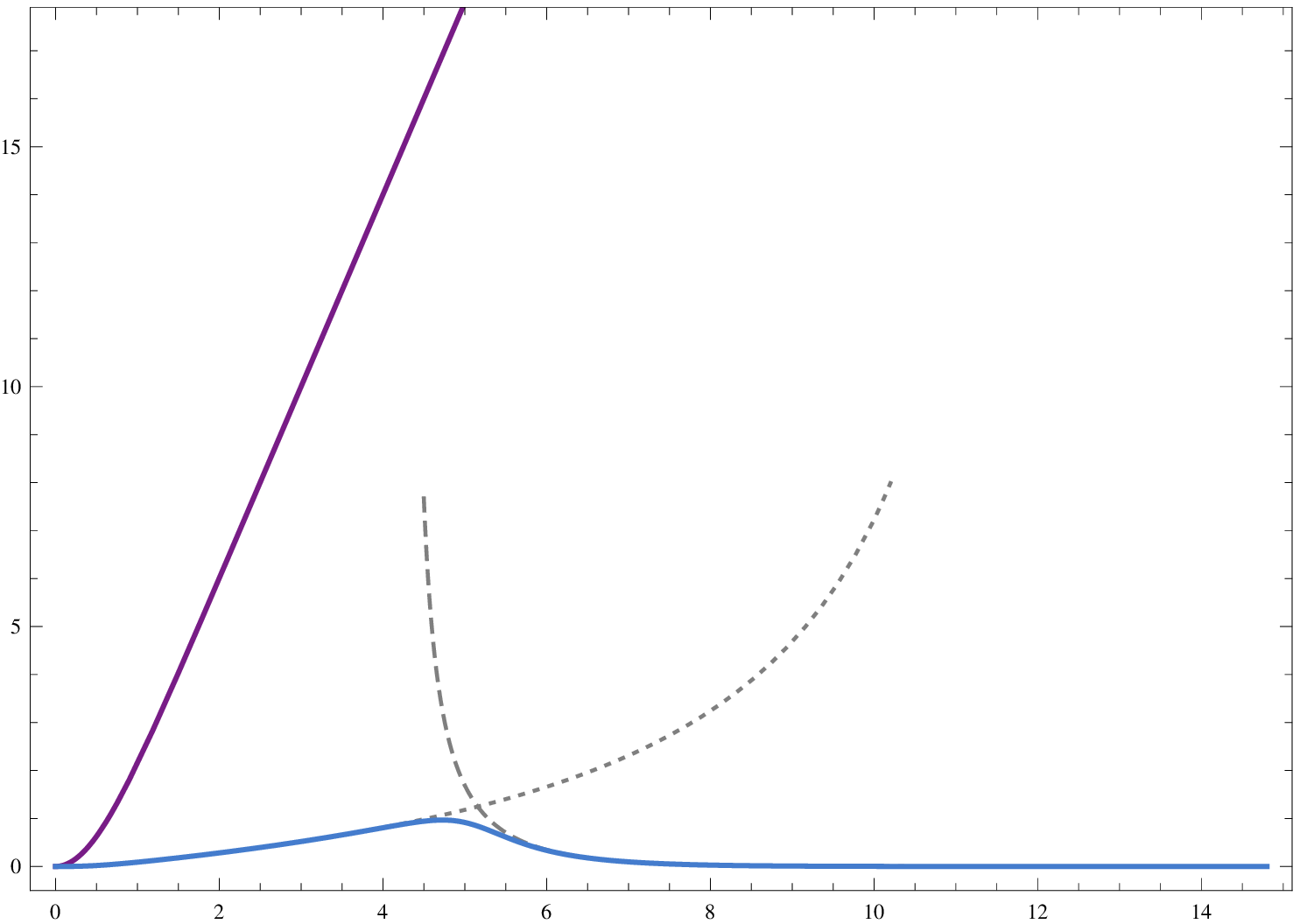}}
\put(160,165){$M_1$}
\put(410,-3){$\r$}
\end{picture} 
\caption{The functions $P$ (left panel), and  $M_1$ (right panel) 
for one example of backgrounds in which $P$ is exponential for large $\r$ (blue),
compared with the $\hat{P}$ solution (purple).}
\label{Fig:plotexpPM1}
\end{center}
\end{figure}

In Fig.~\ref{Fig:plotexpPM1} we show an example of a solution in this class.
As can be seen, $P$ is approximately constant for $\r<\r_{\ast}$, and 
grows exponentially for $\r>\r_{\ast}$. Importantly, $P\gg \hat{P}$,
and as a consequence the baryonic VEV $M_1$ is strongly suppressed with respect to the $\hat{P}$ case.

\begin{figure}[h]
\begin{center}
\begin{picture}(320,170)
\put(-100,0){\includegraphics[height=6cm]{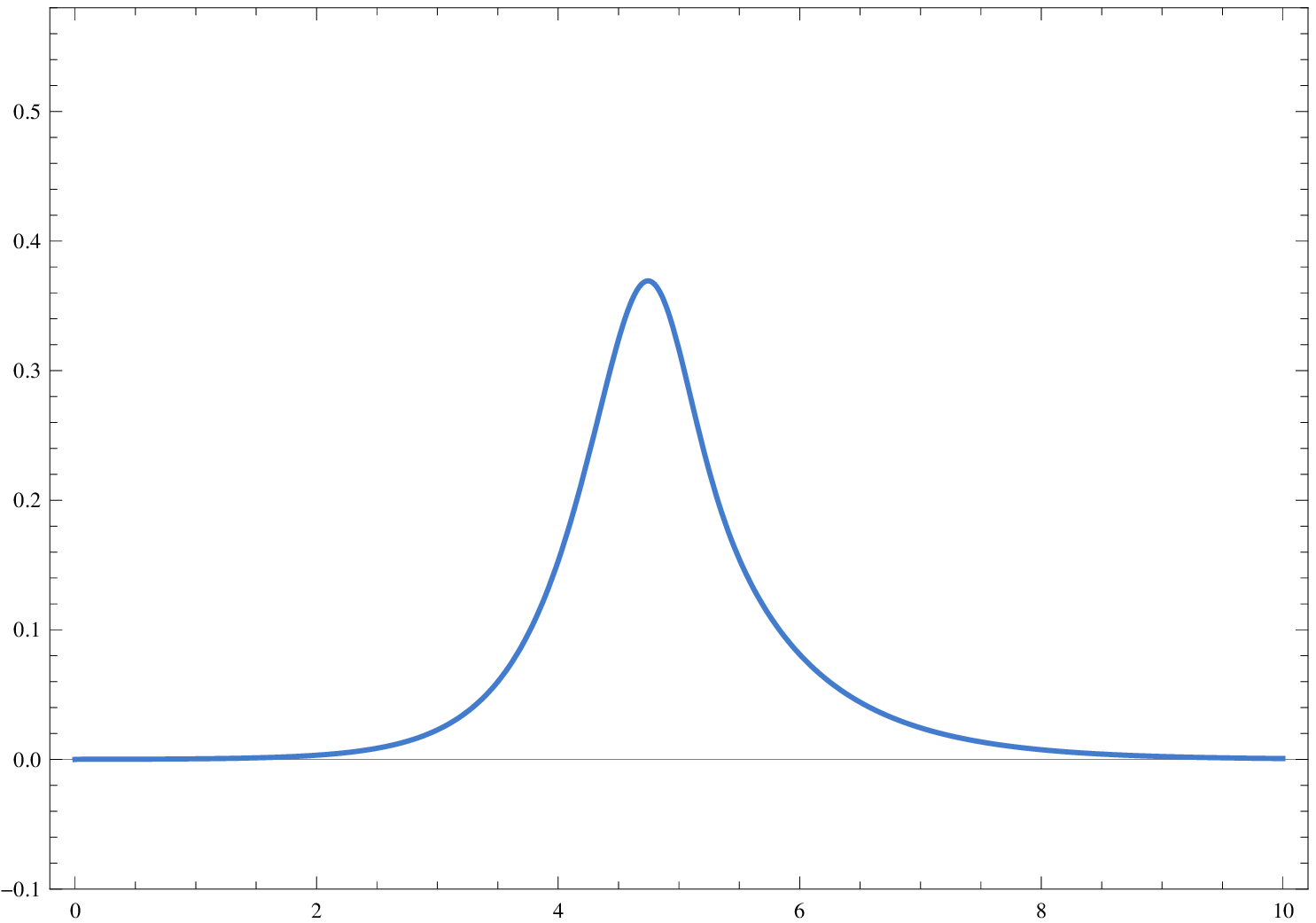}}
\put(-100,162){$\mathcal{Z}$}
\put(143,-3){$\r$}
\put(170,0){\includegraphics[height=6cm]{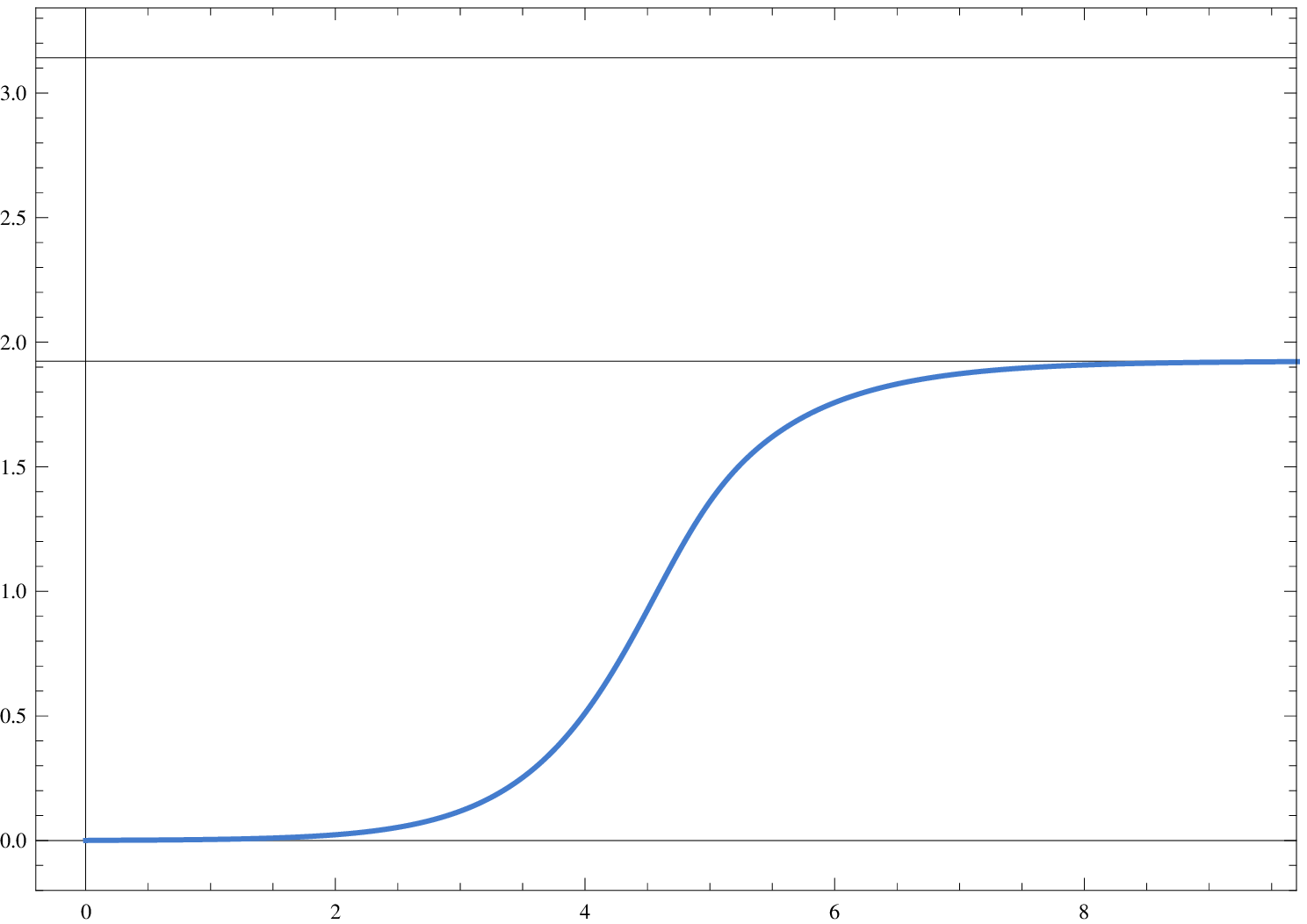}}
\put(167,165){$\bar{\phi}$}
\put(410,-3){$\hat{\r}_0$}
\end{picture} 
\caption{The function $\mathcal{Z}(\r)$ (left panel) and 
 the angular separation $\bar{\phi}$ as a function of the 
turning point $\hat{\r}_o$, for the background with asymptotically exponential $P$ in Fig.~\ref{Fig:plotexpPM1}. Again, the second plot corresponds to an isothermal curve.}
\label{Fig:plotexpzeL}
\end{center}
\end{figure}

\begin{figure}[h]
\begin{center}
\begin{picture}(240,170)
\put(10,3){\includegraphics[height=5.8cm]{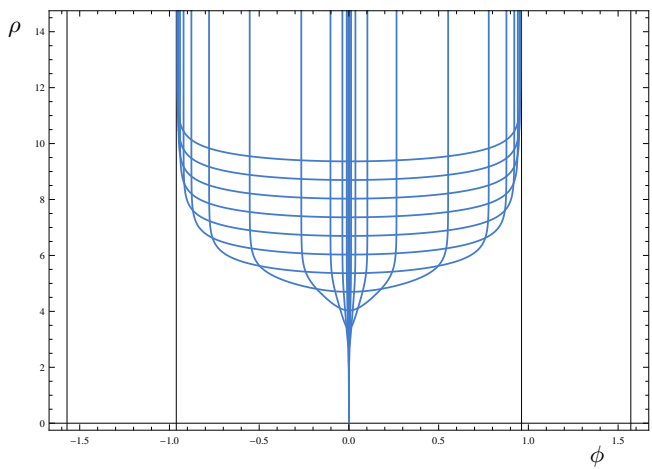}}
\put(0,160){$\r$}
\put(220,-3){${\phi}$}
\end{picture} 
\caption{The $U$-shaped embeddings in the example of background
from Fig.~\ref{Fig:plotexpPM1}, in the plane $(\phi,\r)$, for various choices of $\hat{\r}_o$.}
\label{Fig:plotexpEmb}
\end{center}
\end{figure}

The function ${\cal Z}$ is in this case positive-definite, as can be seen in Fig.~\ref{Fig:plotexpzeL},
which signals a classical instability. Indeed, a tachyon has been found by looking at the 
fluctuations of the embedding~\cite{CLV}. Besides, the asymptotic angular separation $\bar{\phi}$ is 
monotonically decreasing as the probes explore deeper into the bulk. In Fig.~\ref{Fig:plotexpEmb}, we display the shape of the embedding in the $(\r,\phi)$-plane.
As in the case of the walking backgrounds with asymptotically linear $P$,
configurations with $\hat{\r}_o < \r_{\ast}$ develop a non-trivial structure at the turning point, and 
the classical solutions degenerate
into two identical branches with fixed $\phi$ that sit one on top of the other.
Again, we can use this observation to construct a meaningful comparison 
between the energies of the disconnected and connected configurations.

\begin{figure}
\begin{center}
\begin{picture}(60,170)
\put(-80,5){\includegraphics[height=5.8cm]{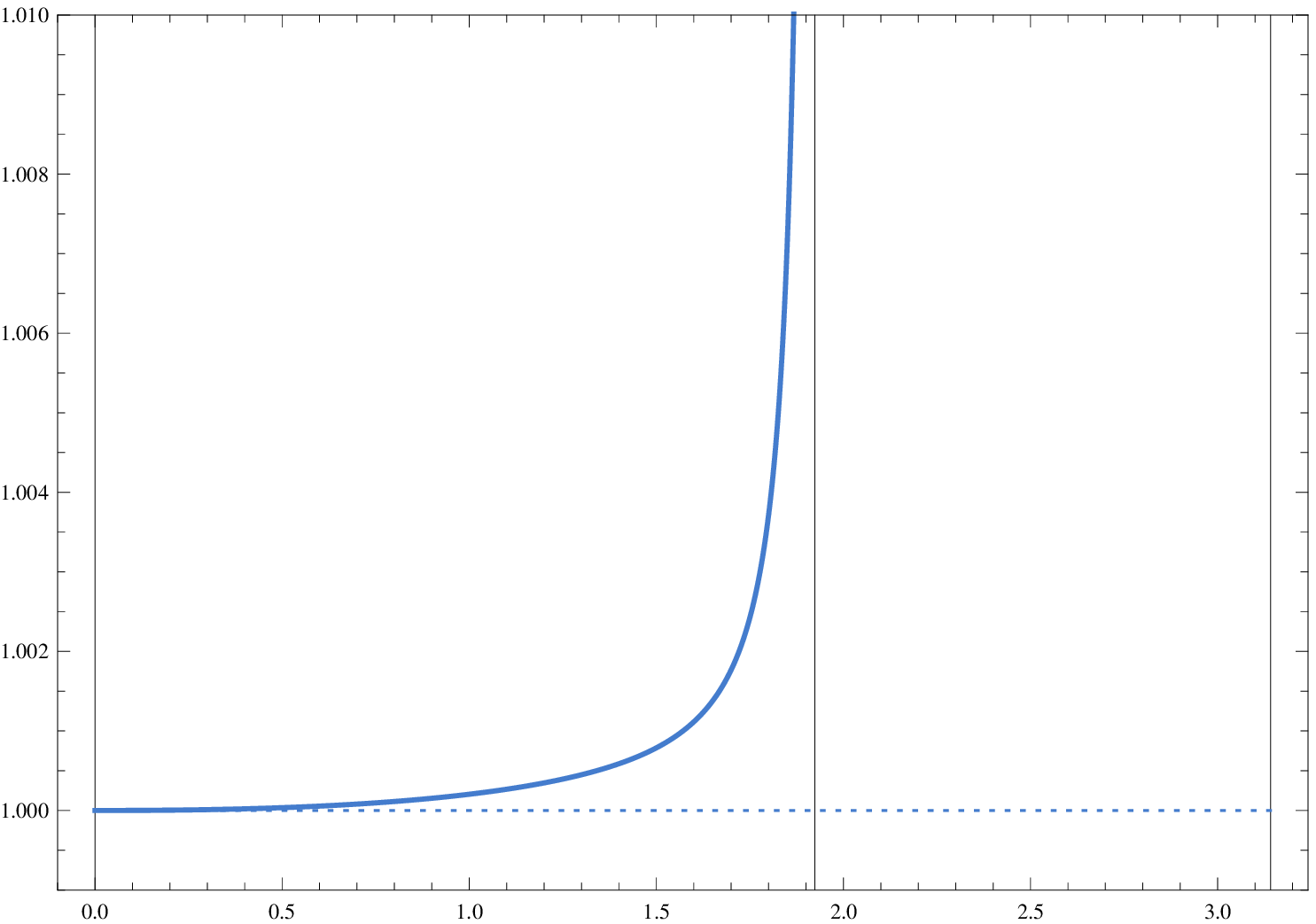}}
\put(-95,155){$E$}
\put(143,0){$\bar{\phi}$}
\end{picture} 
\caption{The energy $E$ as a function of $\bar{\phi}$ for  walking backgrounds 
with asymptotically exponential $P$.  In the plot, the thin dashed lines represent
 the disconnected solutions and the solid thick line represents the connected ones.
  A UV cutoff of $\r=8$ is used when performing the $E$ integral.}
\label{Fig:plotexpwalkEL}
\end{center}
\end{figure}

The result is shown in Fig.~\ref{Fig:plotexpwalkEL}.
As can be seen, the disconnected solution always has lower energy than the connected one.
This is ultimately the reason why the instability found in~\cite{CLV} emerges: 
in backgrounds of this subclass, for any choice of the control parameter $\bar{\phi}$
the disconnected configuration is always energetically favoured, and hence 
the U-shaped embedding is never physically realised.

The conclusion is that this type of U-shaped embedding of probe $D7$ in 
the subclass of walking backgrounds for which 
$P\gg Q$ holds for all $\r$ shows  
fatal pathologies: 
the function ${\cal Z}$ is positive (violating the concavity conditions, and hence yielding a tachyonic 
mode), and furthermore the disconnected configuration is always energetically favoured 
(precluding this construction to be interpreted as a model of chiral symmetry breaking).

We also add a remark about the symmetry properties of these models.
Notice that the condition $P\gg Q$ implies that everywhere in the geometry the $\mathbb{Z}_2$
that exchanges the two $S^2$ in the internal geometry is at least an approximate symmetry.
Comparing this observation with what we found when $P$ is linear (at least asymptotically),
in which case the background severely breaks this $\mathbb{Z}_2$ symmetry, we are tempted to conclude that this
is behind the instabilities we found. As long as the baryonic VEV $M_1$ is non-trivial, the connected solution is predominant dynamically.
But if the background geometry recovers the symmetry, U-shaped embeddings
become unstable in favour of disconnected configurations.
Hence, in this context chiral symmetry breaking cannot be modelled by
the dynamics of embeddings that probe regions of the geometry in which 
the symmetry between the two spheres is present, even approximately.

\section{Conclusions and further directions.}
\label{Sec:conclusions}

In this paper we started by reviewing a large class of Type IIB supergravity solutions
based on the conifold and obtained by wrapping $D5$-branes around an internal two-cycle.
These models can be used to describe the dual of a confining four-dimensional field theory,
in the sense that the standard prescription of gauge/gravity dualities
yields a linear static potential for a
non-dynamical quank-antiquark pair.
We focused our attention on variations of these backgrounds that exhibit walking
behaviour, meaning that the dual gauge coupling 
varies slowly over a finite range of the radial direction, corresponding
to a finite energy interval between two dynamically generated scales.
We reconsidered the proposal in \cite{A} of modelling chiral symmetry breaking
by probing such geometries with a specific type of U-shaped embedding
for $D7$-branes.

Since these embeddings do not preserve supersymmetry, it is pertinent to examine their stability. With this aim, we introduced an efficient diagnostic tool, the function $\mathcal{Z}$ defined in Eq.~(\ref{Zsol}). Under the assumptions of Section~\ref{Sec:extended}, a sufficient condition for perturbative stability is $\mathcal{Z}\le0$. Conversely, the embedding we considered on backgrounds which yield positive $\mathcal{Z}$ (for some range of the radial coordinate ) presents instabilities.
 This requirement can be seen as the result of a concavity condition, similar to the ones encountered for the  thermodynamic potentials. Pursuing this thermodynamic analogy, 
 we also argued that the system will tend to realize the brane configuration that minimizes the energy for a given asymptotic separation of the branes.

In this way, we unveiled the dynamical origin of the instability found in~\cite{CLV}
for the special subclass of models for which the embedding had been originally studied in~\cite{ASW1}
and have asymptotically exponential $P$. According to our analysis, the U-shaped configuration is not a minimum of the action and thus is disfavored with respect to the disconnected, chiral-symmetry preserving solution.

Most importantly, we showed that no such pathologies arise if one considers the same type of embedding on a different subclass of backgrounds (first discussed in~\cite{ENP,NPR}) in which $P$ is asymptotically linear, as in the Maldacena--Nu\~nez solution. Furthermore, we have identified a first-order transition between chiral-symmetry breaking and preserving phases as one increases the asymptotic separation of the branes. The spectrum of the gauge theory dual to this class of solutions contains a parametrically light scalar state~\cite{ENP,EP}, which makes them particularly appealing in the light of the LHC program.

We also commented on the geometric properties of the setup. In particular, we noted that the U-shaped embedding can be realised dynamically and is stable only provided it probes regions of the background in which the $\mathbb{Z}_2$ symmetry exchanging the two $S^2$ factors in the internal geometry is broken. A possible measure of the breaking of this discrete symmetry is the baryonic VEV $M_1$ defined in~Eq.(\ref{barvev}). This quantity reaches its maximum value in the Maldacena-Nu\~nez solution, where the chiral-symmetry breaking phase is dominant over the entire space of parameters. The role of $M_1$ as catalyzer of the breaking, as well as the precise relation (if any) between the geometric $\mathbb{Z}_2$ and the field-theoretic chiral symmetries are intriguing questions. In order to find an answer one would need to better understand how fundamental matter, modelled by the $D7$-branes, couples to the adjoint background. Indeed, it should be possible to write this coupling in terms of $\bar{\phi}$. This is beyond the scope of this paper.

\vspace{1.0cm}
\begin{acknowledgments}
We thank David Mateos for valuable discussions. The work of A. F. was supported by STFC grant ST/J00040X/1, by MEC FPA2010-20807-C02-02, by CPAN CSD2007-00042 Consolider-Ingenio 2010 and finally by ERC Starting Grant ``HoloLHC-306605". The work of M.P. is supported in part by WIMCS and by the STFC grant ST/J000043/1. D. S. is supported by the STFC Doctoral Training Grant ST/I506037/1. 
\end{acknowledgments}


\end{document}